%% file: qgimm16.tex
\documentclass[a4paper,11pt]{article}

\usepackage{amssymb,amstext,amsmath,amsthm}
\usepackage{graphicx}
\usepackage{latexsym}

\usepackage{amsfonts}
\usepackage{verbatim}
\usepackage{color}  

\usepackage{amsmath,amsxtra,amsthm,amssymb,xr}
\usepackage[all]{xy}
\usepackage[]{mathrsfs}

\usepackage{vmargin}

\newcommand{\bea}{\begin{eqnarray}}
\newcommand{\eea}{\end{eqnarray}}
\newcommand{\nn}{\nonumber}
\newcommand{\be}{\begin{equation}}
\newcommand{\ee}{\end{equation}}

\newtheorem{lem}{Lemma}

\newcommand{\R}{\mathbb{R}}
\newcommand{\C}{\mathbb{C}}

\newcommand{\CP}{\mathbb{CP}} 

\DeclareMathOperator{\tr}{tr}

\newcommand{\lalg}[1]{\mathfrak{#1}}  
\newcommand{\SU}{\mathrm{SU}}
\newcommand{\SO}{\mathrm{SO}}
\newcommand{\OO}{\mathrm{O}} 
\newcommand{\SL}{\mathrm{SL}}

\newcommand{\su}{\lalg{su}}
\renewcommand{\sl}{\lalg{sl}}

\newcommand{\so}{\lalg{so}}

\newcommand{\pole}{\mathcal T} 

\newcommand{\dd}{{\mathrm d}}  
\newcommand{\bn}{{\bf n}} 

\def\la{\langle}
\def\ra{\rangle}

\newcommand{\rep}{{\cal H}}  
 \newcommand{\repiso}{{\cal A}} 
\newcommand{\bigJ}{{\cal J}}  
\newcommand{\Rmatrix}{{\cal R}}  
\newcommand{\towerbot}{{\cal I}}  
\newcommand{\SLtwoC}{\SL(2,\mathbb{C})}
\newcommand{\half}{\frac{1}{2}}
\newcommand{\z}{Z}  

\setmarginsrb{2.5cm}{1cm}{2.5cm}{2cm}{12pt}{11mm}{0pt}{13mm}

\title{Lorentzian spin foam amplitudes: graphical calculus and asymptotics}
\author{John W. Barrett$^a$\footnote{john.barrett@nottingham.ac.uk}
, Richard J. Dowdall$^a$\footnote{richard.dowdall@maths.nottingham.ac.uk}
, Winston J. Fairbairn$^a$\footnote{winston.fairbairn@nottingham.ac.uk}, \\
 Frank Hellmann$^a$\footnote{frank.hellmann@maths.nottingham.ac.uk}
, Roberto Pereira$^b$ \footnote{pereira@cpt.univ-mrs.fr}
\vspace{3mm}
\\ $^a$School of Mathematical Sciences \\ Nottingham University \\ University Park \\ Nottingham NG7 2RD \\ UK \\  \vspace{.5mm} \\ $^b$Centre de Physique Th\'eorique \\
Luminy, Case 907 \\ 13288 Marseille - Cedex 9 \\ France}

\date{}

\begin{document}

\maketitle

\vspace{-3mm}
\begin{abstract} The amplitude for the 4-simplex in a spin foam model for quantum gravity is defined using a graphical calculus for the unitary representations of the Lorentz group. The asymptotics of this amplitude are studied in the limit when the representation parameters are large, for various cases of boundary data. It is shown that for boundary data corresponding to a Lorentzian simplex, the  asymptotic formula has two terms, with phase plus or minus the Lorentzian signature Regge action for the 4-simplex geometry, multiplied by an Immirzi parameter. Other cases of boundary data are also considered, including a surprising contribution from Euclidean signature metrics.
\end{abstract}

\section{Introduction}
Spin foam models are discrete versions of a functional integral, usually constructed using a triangulation of the space-time manifold (`foam'), local variables (`spins'), and local amplitudes for the simplexes in the triangulation.
In this paper, a particular spin foam amplitude for Lorentzian signature quantum gravity in constructed and studied in the asymptotic regime. It is shown that, in suitable cases, the asymptotics can be expressed in terms of the action of Regge calculus.

Spin foam models of quantum gravity started with the Ponzano-Regge model for quantum gravity in a three-dimensional space-time with the Euclidean signature of the metric \cite{ponzanoregge}. Four-dimensional spin foam models for quantum gravity were first considered by Reisenberger and Rovelli \cite{Reisenberger:1996pu}, inspired by loop quantum gravity.

Current research in quantum gravity seeks to emulate the success of the Ponzano-Regge model by constructing four-dimensional models which are a direct generalisation of Ponzano-Regge technique. The first such model was constructed by Ooguri \cite{Ooguri:1992eb} from the perspective of matrix models, giving a spin foam model based on flat connections of a gauge field. Models for quantum gravity were
introduced by Barrett and Crane \cite{barrett-1998-39}, first for Euclidean metrics, then for Lorentzian signature metrics \cite{Barrett:1999qw}. Generally, the Euclidean models are easier to define because the representation theory of the rotation group is much simpler than the representation theory of the Lorentz group. The compactness of the rotation group also means that the Euclidean models have fewer regularisation issues.

In the last couple of years a new class of spin foam models for 4d gravity have been introduced \cite{Engle:2007qf,Engle:2007wy,Freidel:2007py,Livine:2007ya}, based on the quantum tetrahedron of Barbieri \cite{Barbieri:1997ks} as the boundary data. These new spin foam models can be seen as an improvement of the Barrett-Crane \cite{barrett-1998-39} model, and incorporate a new parameter which plays the role of the Immirzi parameter in Lagrangian approaches. As usual, the models started in Euclidean versions, the generalisation to the Lorentz group being given first in \cite{Pereira:2007nh}.

In this paper, the approach of \cite{Engle:2007wy} is followed to define and study the amplitude for a 4-simplex in the physical Lorentzian signature quantum gravity. The technical details differ slightly; the approach here being based on a graphical calculus for the representations of the Lorentz group, generalising the `relativistic spin networks' of \cite{Barrett:1999qw}.

It is worth explaining how the new spin foam models differ from the Barrett-Crane model. An important question with the BC model was the relation with a loop quantum gravity \cite{rovelli} boundary geometry, this causing some problems in the computation of the spin foam graviton propagator \cite{Alesci:2007tx,Alesci:2007tg}. This was the main motivation in the work leading to the models proposed in \cite{Engle:2007qf,Engle:2007wy}. This problem is solved in \cite{engle-2007-99,Engle:2007qf,Engle:2007wy} by first introducing the Immirzi parameter $\gamma$ \cite{Immirzi:1996di}\footnote{See \cite{Livine:2001jt} for an earlier discussion on the introduction of the Immirzi parameter in the BC model}, in parallel to its use in LQG and, in a second step, dealing with the `simplicity constraints' of tetrahedron geometry in a different way. In the $\gamma\rightarrow\infty$ limit, one recovers the BC model, while in the $\gamma\rightarrow 0$ limit one recovers the flipped model of \cite{Engle:2007qf}(see \cite{Engle:2007mu} for a discussion on these limits). In a certain sense, and even though the constructions are different, the situation is similar to LQG, where one is able to construct a connection (kinematical) Hilbert space only after introducing the Immirzi parameter. This Hilbert space is generated by $\SU(2)$ spin networks, and it is one of the main results of the work in \cite{Engle:2007qf,Engle:2007wy} that the boundary state space (on a fixed graph) is also spanned by spin networks. An alternative boundary space construction for the BC model, based on projected spin networks, has been proposed in \cite{Livine:2002ak}.

Most of the considerations in the BC paper and in \cite{engle-2007-99,Engle:2007qf,Engle:2007wy} are restricted to a single 4-simplex. Generalizations to larger triangulations, some of them more natural than others, are possible but a certain degree of arbitrariness remains. To resolve this ambiguity, a correct choice of boundary state space is essential. A natural gluing between neighbouring 4-simplices is given by identifying the quantum numbers that specify the boundary states in each simplex, and this was the strategy followed in both the BC model (see for instance \cite{Reisenberger:1998bn}) and in \cite{engle-2007-99,Engle:2007qf,Engle:2007wy}. For the BC model, this leads to a geometrical picture where 4-simplices are glued by identifying the areas of common triangles \cite{Barrett:1997tx}. This picture has the property of leaving some geometrical degrees of freedom uncorrelated, and this fact has drawn some criticism in the literature (see for instance \cite{Livine:2005tp}).

This property of the BC model has also motivated the introduction of closely related models \cite{Freidel:2007py,Livine:2007ya}. There the authors use a coherent state formulation of the boundary space, and these states are used to glue different simplices. For $\gamma<1$ the two constructions of \cite{Engle:2007qf,Engle:2007wy} and \cite{Freidel:2007py,Livine:2007ya} are equivalent, meaning one has both a LQG-like boundary space and an exact gluing of tetrahedron geometries. For $\gamma>1$ the constructions are different -- in \cite{Engle:2007wy} the gluing of geometries is not exact, and for \cite{Freidel:2007py}, the boundary space is spanned by projected spin networks -- but not too different, as explained in \cite{Conrady:2008ea}.

The semiclassical analysis of the models of Euclidean signature for $\gamma<1$ and for arbitrary triangulations of a closed manifold has been studied in \cite{Conrady:2008mk}, where the saddle points for non degenerate configurations are related to Regge-like geometries. The asymptotic analysis of the amplitude for a single 4-simplex was carried out in \cite{Barrett:2009gg}, including the contribution of degenerate configurations and also for any value of $\gamma$ for the Euclidean model of \cite{Engle:2007qf,Engle:2007wy}. The results show that the asymptotics is given in terms of the action of Regge calculus.

The relation with Regge calculus in the semiclassical limit is a remarkable result and gives confidence in these new models as a description of simplicial gravity. It is also a valuable tool in computing observables, e.g., the graviton propagator \cite{Bianchi:2009ri}. In the Euclidean model there are however some puzzling additional terms (the `weird terms') in the asymptotics which have the same Regge calculus phase factor but with the Immirzi parameter absent. A priori it seems that there should be interference between these terms and the terms with the Immirzi parameter; the physical interpretation of this is unclear.

The results of this paper are as follows. There is a precise definition of the spin network calculus for the unitary representations of the Lorentz group, leading to a definition of the 4-simplex amplitude along similar lines to \cite{Pereira:2007nh,Engle:2007wy}, but using bilinear inner products instead of Hermitian ones.
 Then the stationary phase method is applied to the resulting integrals to give an asymptotic formula for the amplitude in the limit where the representation parameters are all large. The geometry of the critical points is analysed, giving a formula for the phase part of each contribution to the asymptotics in terms of the Regge action for a geometric simplex determined by the boundary data, following the method developed in \cite{Barrett:2009gg}. For boundary data which determines a Lorentzian signature metric inside the 4-simplex, the action is the Lorentzian Regge action, as one would hope. In contrast to the Euclidean case, there are no additional `weird terms'. Instead, these reappear on their own in the case of Euclidean boundary data. Some further remarks about the asymptotic results are given in section \ref{conclusions}.

\section{Graphical calculus for the Lorentz group}

Throughout this paper, $\SL(2,\C)$ refers to the 6-dimensional real Lie group of $2\times2$ complex matrices with unit determinant, and is called simply the Lorentz group. It covers the group of proper orthochronous Lorentz transformations, $\SO^+(3,1)$, which is the component of the group $O(3,1)$ connected to the identity.


\subsection{Representations}

The principal series of irreducible unitary representations of the Lorentz group $\SL(2,\C)$ are labelled by two parameters $(k,p)$, with $k$ a half-integer and $p$ a real number \cite{GGV}. The representations are in a Hilbert space $\rep_{(k,p)}$, the labels $(k,p)$ indicating the action of the group in this space. The Hermitian inner product is denoted $(\psi,\chi)$ for vectors $\psi,\chi\in\rep_{(k,p)}$.

Some standard facts about these representations are as follows; explicit formulae with which one can check these results are given a little later.

\begin{itemize}\item The $(k,p)$ representation splits into the irreducible representations $V_j$ of the $\SU(2)$ subgroup as
$$\rep_{(k,p)}=\bigoplus_{j=|k|}^\infty V_j$$
with $j$ increasing in steps of 1.
\item There is a unitary intertwiner of representations
$$\repiso\colon \rep_{(k,p)}\to \rep_{(-k,-p)}$$
\item There is an anti-linear map
$$\bigJ\colon \rep_{(k,p)}\to \rep_{(k,p)}$$
which commutes with the group action, and is unitary in the sense that $(\bigJ\psi,\bigJ\chi)=(\chi,\psi)$. It satisfies $\bigJ^2=(-1)^{2k}$.
\item There is an invariant bilinear form $\beta$ on $\rep_{(k,p)}$ defined by
$$\beta(\psi,\chi)=(\bigJ\psi,\chi).$$
The properties of $\bigJ$ show immediately that $\beta(\psi,\chi)=(-1)^{2k}\beta(\chi,\psi)$, so that $\beta$ is symmetric or antisymmetric as $2k$ is even or odd.
\end{itemize}

Both $\repiso$ and $\bigJ$ respect the decomposition into $\SU(2)$ subspaces; clearly $\repiso$ restricts to a multiple of the identity operator on the $j$-th subspace, whilst $\bigJ$ restricts to a phase times the standard antilinear map $J$ on an $\SU(2)$ representation. Explicit formulae fixing these constants are given below.

These elements are almost all one needs to start constructing a diagrammatic calculus along the lines of the spin network calculus for $\SU(2)$. The one remaining piece of data needed is the R-matrix
$$\Rmatrix\colon \rep_{(k,p)}\otimes \rep_{(k',p')}\to \rep_{(k',p')}\otimes \rep_{(k,p)}$$
given by $\Rmatrix\colon x\otimes y\to (-1)^{4kk'}y\otimes x$.

 \subsection{Lorentzian diagrammatic calculus}

The idea of a diagrammatic calculus is to organize calculations in tensor calculus according to a diagram drawn in the plane. As the Lorentz group is non-compact, some of the constructions routinely used for a compact group will not always be well-defined. Accordingly, the diagrammatic calculus for the Lorentz group presented here is limited to the description of a certain class of  diagrams. The elements of the diagrammatic calculus for the Lorentz group are as follows.

\subsubsection{Lorentzian spin networks}

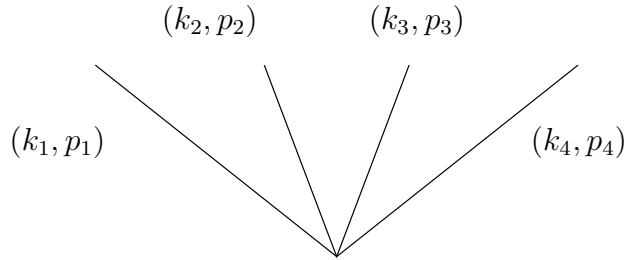
\begin{figure}[htb]
\begin{center}
\input{vertex.pstex_t}
\caption{A spin network vertex}\label{vertex}
\end{center}
\end{figure}

An element of the tensor product of vector spaces $\tau\in\rep_{(k_1,p_1)}\otimes\ldots\otimes\rep_{(k_n,p_n)}$ is represented diagrammatically by a vertex labelled with $\tau$ having $n$ `legs' pointing upwards, labelled with $(k_1,p_1)$ to $(k_n,p_n)$ from left to right (figure \ref{vertex}). Note that the tensor $\tau$ is not required to be an invariant tensor. For every vertex it is required that $k_1+\ldots+k_n$ is an integer. This requirement will be important later to show that vertices can be permuted. It is also a requirement for non-zero tensors invariant under the $\SU(2)$ subgroup of $\SL(2,\C)$, which will also be important later.

The tensor product of several such tensors $\tau_1\otimes\tau_2\otimes\ldots$ is represented by the diagram formed by placing the vertices in a horizontal line, in the order in which they appear in the tensor product.
Scalars are formed from these tensors by contracting pairs of legs labelled with the same $(k,p)$ using the bilinear form $\beta$.  The precise details of this contraction are organised with the aid of a diagram.

\begin{figure}[htb]
\begin{center}
\input{maximum.pstex_t}
\caption{The maximum diagram}\label{maximum}
\end{center}
\end{figure}
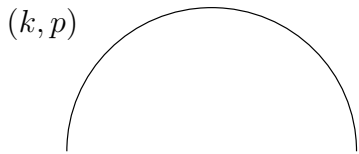

The bilinear form $\beta$ for the representation $\rep_{(k,p)}$ is represented by a `maximum' diagram labelled with $(k,p)$ (figure \ref{maximum}). Since (for odd half-integer $k$) the order of the two arguments of the bilinear form is important, the convention is that the first argument corresponds to the left-hand end of the line, and the second argument to the right-hand end. In this diagrammatic calculus there is however no minimum diagram.

\begin{figure}[htb]
\begin{center}
\input{crossing.pstex_t}
\caption{The crossing diagram}\label{crossing}
\end{center}
\end{figure}
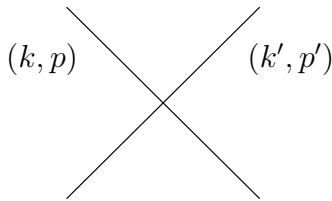

The crossing diagram (figure \ref{crossing}) stands for the linear map $\Rmatrix$, and a line that has no minima, maxima or crossings (a `vertical' line) is the identity map.

\begin{figure}[htb]
\begin{center}
\input{RImove.pstex_t}
\caption{Reidemeister move I}\label{ReidemeisterI}
\end{center}
\end{figure}
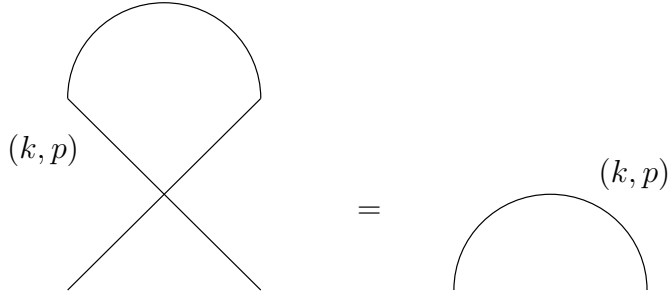

The diagrams are composed horizontally by tensor product, and vertically by the composition of maps. For example, the crucial property relating $\Rmatrix$ and $\beta$ is
\begin{equation}\beta\circ\Rmatrix=\beta,\label{RI}\end{equation}
 which is represented diagrammatically in figure \ref{ReidemeisterI}.

This property explains the choice of the R-matrix, and is analogous to the binor calculus for $\SU(2)$  \cite{penrose,Barrett:2008wh}. In fact this R-matrix reduces to the $\SU(2)$ one for the binor calculus on the $V_j$ subspaces of each $\rep_{(k,p)}$.

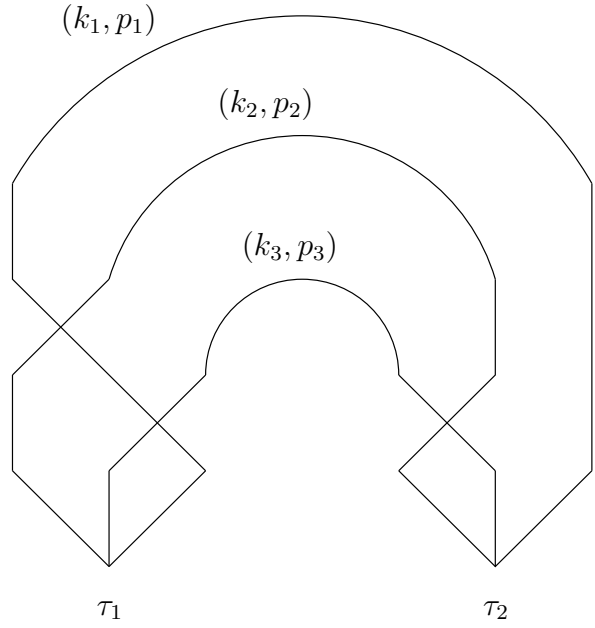
\begin{figure}[htb]
\begin{center}
\input{spinnetwork.pstex_t}
\caption{An example of a Lorentzian spin network}\label{spinnetwork}
\end{center}
\end{figure}

The diagram for a scalar consists of a set of vertices, arranged in a horizontal line at the bottom of the diagram, composed with the diagrammatic elements introduced above: crossings, vertical lines, and semi-circular maximum diagrams, so that there are no remaining free ends. An example is shown in figure \ref{spinnetwork}.

The value of the diagram is obtained by composing all of the corresponding tensors and linear maps to give a number. Abstractly, this can be viewed as applying a monoidal functor from the diagram category generated by the vertices, crossings and maxima to vector spaces.
In this composition, the R-matrices contribute the overall sign
\begin{equation}\prod_\text{crossings}(-1)^{4kk'}.\label{signfactor}\end{equation}
For example, for figure \ref{spinnetwork}, if $\tau_1=A\otimes B\otimes C$ and $\tau_2=D\otimes E\otimes F$, then the value of the diagram is
$$(-1)^{4(k_1k_3+k_2k_3+k_1k_2)}\beta(A,D)\beta(B,E)\beta(C,F).$$

\begin{figure}[htb]
\begin{center}
\input{spinnetwork2.pstex_t}
\caption{An alternative diagram}\label{spinnetwork2}
\end{center}
\end{figure}
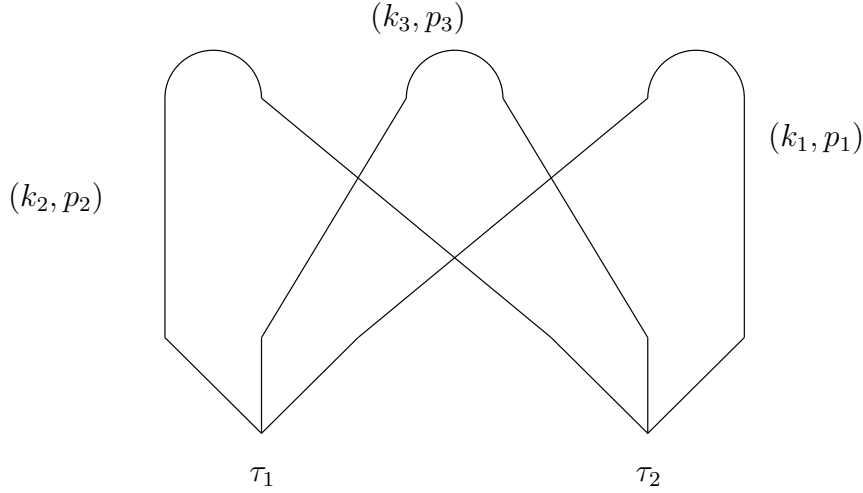

There are many diagrams which connect together the same legs in pairs.  For example, a diagram can be drawn so that it represents
\begin{equation}(\beta\otimes\beta\otimes\ldots\otimes\beta)\circ P \circ (\tau_1\otimes\tau_2\otimes\ldots)\label{Wick}\end{equation}
 with $P$ the diagram of a permutation, consisting of a number of instances of $\Rmatrix$ and the identity composed together both horizontally and vertically. Such a diagram with the same connectivity as figure \ref{spinnetwork} is given in figure \ref{spinnetwork2}.

The power of the diagrammatic method lies in the fact that however the closed diagram is drawn according to the above rules, the resulting value is always the same. This means that given the following combinatorial data
\begin{itemize}
\item A graph $\Gamma$, i.e. a collection of vertices and edges joining the vertices
\item For each vertex of $\Gamma$, an ordering of the edges meeting the vertex
\item For each edge of $\Gamma$, a representation $(k,p)$
\item For each vertex of  $\Gamma$, a tensor $\tau$ in the vector space determined by the representations and ordering at the vertex
\end{itemize}
then every diagram constructed from this data has the same value.

The argument to prove this is as follows. Construct a diagram drawn in the plane according to the rules given above with the vertices arranged in some particular order from left to right along the bottom of the diagram. Consider two distinct edges of the graph, $M$ and $N$.
In the diagram, $M$ and $N$ cross an even number of times if both ends of $M$ lie between the ends of $N$ or both ends of $N$ lie between the ends of $M$. Otherwise, they cross an odd number of times. Thus the part of the sign factor \eqref{signfactor} due to the crossing of $M$ and $N$ does not depend on how the diagram is drawn. This is true for all distinct $M$ and $N$. Therefore the only  signs in \eqref{signfactor} not yet accounted for are due to the self-crossing of an edge. However due to \eqref{RI}, the combination of the maximum diagram and a number of self-crossings always yields the same inner product $\beta$.

\begin{figure}[htb]
\begin{center}
\input{vertexperm.pstex_t}
\caption{Permuting the vertices in a diagram}\label{vertexperm}
\end{center}
\end{figure}
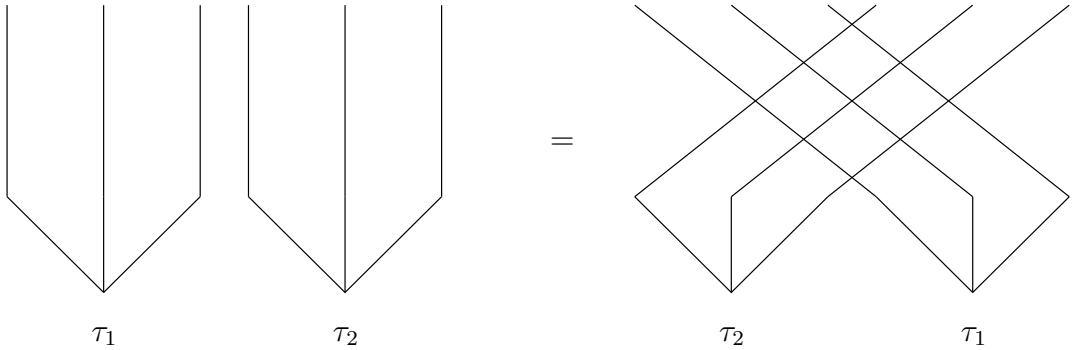

Finally, consider changing the order of the vertices in a closed diagram by permuting two neighbouring vertices. A new diagram can be constructed by replacing the two vertices on the left of figure \ref{vertexperm} by the right-hand side, thus giving a closed diagram for the same combinatorial data but with the vertices in the diagram permuted.
However the tensor evaluation of the two sides of figure \ref{vertexperm} are equal. This is because the
right-hand side can be written as $Q(\tau_2\otimes\tau_1)$, and the
 sign factor in the permutation $Q$ is equal to 1, as follows from the fact that $\sum k$ is an integer for each vertex. Therefore
$$\tau_1\otimes\tau_2=Q(\tau_2\otimes\tau_1),$$
and hence the evaluation of a closed diagram is independent of the ordering of the tensors along the horizontal axis. This completes the proof.

It is worth noting that the evaluation of \eqref{Wick}, including the signs, is exactly the one obtained for a Feynman diagram according to Wick's theorem if the legs with integer $k$ are treated as bosons and half-integer $k$ fermions, and the vertices as interactions. Accordingly, the inner product $\beta$ is called the propagator.

 The above proof would of course be trivial if all of the $k$ variables were integers, for then the propagators are all symmetric and the crossing factors all $1$. Thus the evaluation is obviously determined by the combinatorial structure alone. However in the fermionic case this is not trivial. Just using the structure of the graph $\Gamma$, it would not be clear which way round to use an antisymmetric propagator.

The dependence on the ordering of the edges at each vertex can not be removed in general, for this is needed to define the tensor product. However the permutation diagrams give an action of the permutation group on the vertex tensors, so the tensors for different choices of ordering of the legs at a vertex are canonically related.

\subsubsection{Comparison with the compact case}

The definitions given above have analogues for the representations of the compact group $G=\SU(2)\times\SU(2)$, or just $G=\SU(2)$. However in those cases one also has a definition of the inverse of $\beta$ as a tensor, which is represented as a minimum diagram. Then the diagrams can be any plane projections of graphs embedded in $\R^3$ \cite{RT}. In the case of the Lorentz group, the inverse of $\beta$ would be a tensor of infinite norm, so does not fit within the framework constructed here. This problem manifests itself in the fact that the unknot is a valid diagram containing a minimum, but its value should be a sign times the dimension of the representation, which is infinite.

 A second major difference is that in the compact cases one usually works with only intertwining operators at the vertices. This is important for the geometrical interpretation of the construction: if the representations stand for quantized bivectors, then the invariant tensors with four legs stand for a quantized set of bivectors which sum to zero, as is the case for the bivectors belonging to the faces of a tetrahedron in four-dimensional Euclidean space.

In the present framework that would mean using invariant tensors in $\rep_{(k_1,p_1)}\ldots\rep_{(k_n,p_n)}$. However non-zero invariant tensors do not exist for the Lorentz group, due to the infinite-dimensional nature of the Hilbert spaces\footnote{Using two such invariant tensors and the inner product, one could construct a non-zero invariant operator $\rep_{(k_1,p_1)}\to\rep_{(k_1,p_1)}$ with finite trace, which is impossible.}.
To get around this problem, the diagrammatic calculus is first constructed using vectors $\tau$ which are not invariant\footnote{The fact that non-invariant tensors should be relevant in the construction of a spin foam model for gravity has been pointed out by Alexandrov \cite{Alexandrov:2007pq}.}, and then an averaging procedure is used to define, in favourable cases, Lorentz-invariant spin networks.

\subsubsection{Lorentz invariance}
 The action of the Lorentz group on each of the tensors at vertices is considered by calculating the diagram evaluation using $X\tau$, with $X\in\SL(2,\C)$. In the compact case one would average over the group $G$ at each vertex to get an invariant tensor
$$\iota=\int_{X\in G} X\tau $$
However for the Lorentz group there is no way of normalising this integral to form an average, and using the standard unnormalised integral would result in a vector of infinite norm. The solution to this problem, as in \cite{Barrett:1999qw}, is to write a formal integral for the entire diagram and regularise it by dropping one of the integrals.  In fact, after performing all but one of the integrals, the amplitude is independent of the one remaining $X$ variable.

This procedure was investigated in detail for the Lorentz group representations with $k=0$ in \cite{Baez:2001fh,Cherrington:2005mz,Christensen:2005tr}, where it was found that although the regularised integrals do not always converge, the ones of interest for the state sum model do. Similar results are obtained in the generalisation to arbitrary $k$ presented here \cite{Engle:2008ev}.

\subsection{Explicit formulae}

The standard representation is to take  $\rep_{(k,p)}$ to be the space of functions of two complex variables $z=(z_0,z_1)$ that are homogeneous
$$f(\lambda z)=\lambda^{-1+ip+k}{\bar\lambda}^{-1+ip-k}f(z).$$
The inner product on this vector space is defined using the standard invariant 2-form on $\C^2-\{0\}$ given by
$$\Omega=\frac i2(z_0\dd z_1-z_1\dd z_0)\wedge (\bar z_0\dd \bar z_1-\bar z_1\dd \bar z_0).$$
For $f,g\in \rep_{(k,p)}$, the form $\bar f g\Omega$ has the right homogeneity to project down to $\CP^1$.
The inner product is given by integrating this 2-form.
$$(f,g)=\int_{\CP^1}\bar f\, g\; \Omega.$$
Alternatively, one can do the integration on a section of the bundle $\C^2-\{0\}\to\CP^1$. Gelfand's choice is the section $\zeta\mapsto (\zeta,1)$, for which the integration measure reduces to the standard measure on the plane, $\Omega=\frac i2 \dd \zeta\wedge \bar \dd \zeta=\dd x\wedge\dd y$, with $\zeta=x+iy$.

The element $X\in\SL(2,\C)$ acts on a homogeneous function by
\begin{equation}\label{action}(Xf)(z)=f(X^Tz),\end{equation}
which uses the transpose matrix $X^T$. This gives the unitary representation in the principal series.

Our standard notation for structures on $\C^2$ is as follows.
The Hermitian inner product is
$$\la z,w\ra =\bar z_0w_0 + \bar z_1w_1,$$
the $\SU(2)$ structure map
$$J\begin{pmatrix}z_0\\z_1\end{pmatrix}=\begin{pmatrix}-\bar z_1\\\bar z_0\end{pmatrix},$$
and the antisymmetric bilinear form is
$$[z,w]=z_0w_1-z_1w_0= \la Jz,w\ra.$$

The unitary isomorphism $\repiso$ for $(k,p)\ne(0,0)$ is a scalar multiple of the map defined in \cite{GGV}
$$\repiso f(w)= c'\int_{\CP^1} [w,z]^{-1-k-ip}\overline{[w,z]}^{-1+k-ip} f(z)\;\Omega$$
with the constant $c'=\sqrt{k^2+p^2}/\pi$.

There is also an anti-unitary isomorphism $\rep_{(k,p)}\to \rep_{(-k,-p)}$
given by complex conjugation of functions. Combining these two gives the antilinear stucture map
$$\bigJ f=\overline{\repiso f}.$$

According to \cite{GGV},
$$\repiso\repiso f=(-1)^{2k} f,$$
mapping from $\rep_{(k,p)}$ to $\rep_{(-k,-p)}$ and back.
A short calculation shows that this implies also $\bigJ^2=(-1)^{2k}$, this time both mappings being on the same space. These relations will be verified again explicitly below, using coherent states.

Using these definitions, the bilinear form is
\be\label{bilinear}\beta(\psi,\chi)= c'\int_{\CP^1\times\CP^1} [w,z]^{-1-k-ip}\overline{[w,z]}^{-1+k-ip} \psi(z)\chi(w)\;\Omega_z\;\Omega_w.\ee
These integrals are a little tricky since each power of $[z,w]$ does not separately exist as a function on the whole plane; one has to combine them. Writing $[w,z]=re^{i\theta}$, the integrand contains
$$r^{-2-2ip}e^{-2ik\theta},$$
which is well-defined. It is also clear from this formula that $\beta(\psi,\chi)=(-1)^{2k}\beta(\chi,\psi)$.

\section{The spin foam amplitude}

In this section, the formalism presented above is used to construct $4$-simplex amplitudes for spin foam models of Lorentzian quantum gravity. We follow the Lorentzian EPRL construction \cite{Engle:2007wy} in the choice of intertwining operators but keep the representation labels general and use bilinear forms instead of Hermitian inner products for the construction. This last point suppresses ambiguities in the definition of the corresponding spin networks.

\subsection{Boosted $\SU(2)$ intertwiner}

A four-simplex geometry has as its boundary five geometrical tetrahedra. For the construction of the four-simplex amplitude in this paper, it is assumed that these tetrahedra are non-degenerate and spacelike. Accordingly, they have a Euclidean geometry. The quantization of such a Euclidean geometry is the quantum tetrahedron of Barbieri \cite{Barbieri:1997ks}, in which each triangle is assigned an irreducible representation of $\SU(2)$, and the tetrahedron an intertwiner (invariant vector) in the tensor product of the four representations of its triangles
$$\hat\iota\in V_{k_1}\otimes V_{k_2}\otimes V_{k_3}\otimes V_{k_4}.$$

The four-simplex amplitude however requires the specification of two parameters $(k,p)$ for each triangle, specifying an irreducible representation of the Lorentz group. Since $\rep_{(k,p)}$ is isomorphic to $\rep_{(-k,-p)}$ it is not necessary to consider both of these representations. Therefore, it is assumed that $k\ge0$.

The main assumption of \cite{Engle:2007wy} is that this parameter $k$ is the same as the one specifying the $\SU(2)$ representation on the boundary. Then the $\SU(2)$ representation is isomorphic to the lowest member of the tower of $\SU(2)$ representations contained in $\rep_{(k,p)}$.

Hence the construction \cite{Engle:2007wy} uses a fixed unitary injection
$$\towerbot\colon V_k\to \rep_{(k,p)},$$
for each $k$ and $p$, which commutes with $\SU(2)$. An explicit formula for this injection is defined as follows. The vector space $V_k$ can be defined as the space of homogeneous polynomials in $z=(z_0,z_1)$ of degree $2k$, with the action of $\SU(2)$ the same formula as \eqref{action}. The injection is defined by taking a polynomial $\phi(z)$ to
$$\towerbot\phi(z)=\la z,z\ra^{-1+ip-k}\phi(z).$$
A homogeneous polynomial can be written explicitly as
$$\phi(z)=\phi^{a_1a_2\ldots a_{2k}}z_{a_1}z_{a_2}\ldots z_{a_{2k}}.$$
A calculation shows that the inner product on $V_k$ which makes $\towerbot$ unitary is the one which satisfies
$$|\phi|^2=\frac\pi{2k+1}\bar\phi^{a_1a_2\ldots a_{2k}}\phi^{b_1b_2\ldots b_{2k}}
\delta_{a_1b_1}\delta_{a_2b_2}\ldots\delta_{a_k b_k}.$$
Applying $\towerbot$ on each factor then gives an injection
$$ V_{k_1}\otimes V_{k_2}\otimes V_{k_3}\otimes V_{k_4}\to
\rep_{(k_1,p_1)}\otimes\rep_{(k_2,p_2)}\otimes\rep_{(k_3,p_3)}\otimes\rep_{(k_4,p_4)}.$$
 For convenience, this map will also be denoted by $\towerbot$.

The quantum tetrahedron then gives a tensor
$\towerbot \bigl(\hat\iota\bigr)$, which is not Lorentz invariant. Acting on this by the complex conjugate\footnote{The reason why we are acting with the complex conjugate matrix $\bar{X}$ and not $X$ is simply for the convenience of the notation in what follows.} of a Lorentz transformation $X\in\SL(2,\C)$ then gives the vertex factor
$$\bar{X}  \towerbot \bigl(\hat\iota\bigr)$$
corresponding to the tetrahedron.

In the original EPRL construction, the representation labels are restricted to those such that $ p = \gamma k$, where the constant $\gamma \in \R$ is the Immirzi parameter, in order to implement the full simplicity constraints. Here, the labels are initially left as general as possible. Interestingly, we will see in the asymptotic analysis that only the labels for which $p$ and $k$ are proportional admit critical points, and that labels which do not satisfy a relation of this type are suppressed.

\subsection{Definition of the amplitude}

The 4-simplex amplitude is evaluated using the graph $\Gamma$ of edges of the dual complex to the boundary of the 4-simplex. This graph has a vertex at the centre of each tetrahedron and an edge dual to each triangle of the 4-simplex.

Labelling the tetrahedra of the boundary of the simplex with the latin indices $a,b=1,...,5$, the input data for the 4-simplex amplitude are an irreducible $(k_{ab},p_{ab})$ of the Lorentz group labelling the triangle common to the $a$-th and the $b$-th tetrahedra, and an intertwiner $\hat\iota_a$ on the $a$-tetrahedron. To specify this intertwiner unambiguously, it is necessary to give an order of the four triangles in each tetrahedron. This can be done in any convenient way.

A complex number $\la\Gamma\ra$ is  defined by evaluating the Lorentzian spin network $\Gamma$ with the edges labelled with the corresponding $(k_{ab},p_{ab})$ and the vertices with $\bar X_a  \towerbot \bigl(\hat\iota_a\bigr)$, using the given ordering for its legs. This evaluation is a function of the five variables $X_a\in\SL(2,\C)$. Finally the 4-simplex amplitude is defined by
$$f_4=\int_{\SL(2,\C)^4} \la\Gamma\ra\; \dd X_1 \dd X_2\;\dd X_3\;\dd X_4.$$
The integral over the $\SL(2,\C)$ variables is performed using the Haar measure and is regularised by omitting one of the integrals, the integral over $X_5$, exactly as in \cite{Barrett:1999qw}. The formula for $f_4$ is in fact independent of $X_5$, so integrating over it would certainly give an infinite answer.

\subsection{Coherent states and propagators}

The boundary states used for the asymptotic analysis are those determined by coherent states in each $\SU(2)$ representation \cite{Barrett:2009gg}. These states have the property that a definite geometry emerges for each tetrahedron in the semiclassical limit \cite{livine-2007-76}.

A coherent state \cite{perelomov} is determined by a spinor $\xi\in\C^2$. The coherent state in $V_k$ is the tensor product of $2k$ copies of $\xi$, and is an eigenvector of the spin operator, with maximal weight, along the axis in three-dimensional space determined by $\xi$. Coherent states satisfy a number of interesting properties that will be used in the sequel. Firstly, to a coherent state $\xi$ in $\C^2$ is associated a unit vector $\mathbf{n}_{\xi}$ in $\R^3$, where we have introduced a bold letter notation for $3$-vectors. To the coherent state $J \xi$ is associated the antipodal vector $-\mathbf{n}_{\xi}$. These coherent states are determined up to a phase, that turns out to play an important geometrical role, and they are normalised, i.e., $\langle \xi , \xi \rangle = 1$.

As a polynomial, a normalised coherent state is written
$$\phi(z)= \sqrt{  \frac{d_k }{\pi}} [z,\xi]^{2k}    ,$$
with $d_k = 2k + 1$. This gives
$$\towerbot\phi(z)= \sqrt{  \frac{d_k }{\pi}}       \la z,z\ra^{-1+ip-k}[z,\xi]^{2k}   $$
in $\rep_{(k,p)}$.

A calculation of the integral shows that for $(k,p)\ne(0,0)$
\be
\label{Jformula}
\bigJ\towerbot\phi(z)=(-1)^{2k} \sqrt{  \frac{d_k }{\pi}}        \frac{\sqrt{k^2+p^2}}{k+ip}\la z,z\ra^{-1+ip-k}[z,J\xi]^{2k},
\ee
from which one can again verify $\bigJ^2=J^2=(-1)^{2k}$.
This calculation is done as follows.
We calculate the action of $\repiso$ on $\towerbot\phi(w)$ by also multiplying by $ \overline {\la w,\xi \ra}^{2k}$
\bea
\overline {\la w,\xi \ra}^{2k}   \repiso \towerbot \phi(w)
&=& c' \sqrt{  \frac{d_k }{\pi}} \int_{\CP^1}
\la z,z\ra^{-1+ip-k}
\la J w,z\ra^{-1-k-ip}
\la z ,Jw \ra^{-1+k-ip}
\la \xi ,w \ra^{2k}
\la J z,\xi \ra^{2k}\;\Omega_z
\nn \\
&=&
(-1)^{2k} c'  \sqrt{  \frac{d_k }{\pi}}\int_{\CP^1}
\la z,z\ra^{-1+ip-k}
(\la J w,z\ra
\la z ,Jw \ra)^{-1-k-ip}
\;\times \nn\\ &&\times\;
(\la \xi ,w \ra
\la w ,Jz \ra
\la J z,\xi \ra)^{2k}\;\Omega_z
\nn
\eea
Where we have used $J^2 = (-1)^{2k}$. We can simplify the integration by setting $|z|=|w|=1$, and now take the integral over the 2-sphere $\mathcal{S}^2$. We also use the following identity for unit spinors
\be
\label{null}
z \otimes z^{\dagger} = \frac{1}{2}(1 + {\pmb \sigma} \cdot \mathbf{n}_z).
\ee
Here, ${\pmb \sigma} = (\sigma_1, \sigma_2, \sigma_3)$ are the Hermitian Pauli matrices with eigenvalues $\pm 1$ (see Appendix B), and the dot $\cdot$ is the $3d$ Euclidean scalar product. The symbol $\dagger$ is for Hermitian conjugation. This gives
\bea
\overline {\la w,\xi \ra}^{2k}   \repiso \towerbot \phi(w)
&=&  (-1)^{2k} c' \sqrt{  \frac{d_k }{\pi}} \pi \int_{\mathcal{S}^2}
\tr \left(
\frac{1}{8}(1 + {\pmb \sigma} \cdot \mathbf{n}_{\xi})
(1 + {\pmb \sigma} \cdot \mathbf{n}_{w})
(1 - {\pmb \sigma} \cdot \mathbf{n}_{z})        \right)^{2k}
\nn \\
&\times&
\tr \left(
\frac{1}{4}(1 + {\pmb \sigma} \cdot \mathbf{n}_{z})
(1 - {\pmb \sigma} \cdot \mathbf{n}_{w})
                                                        \right)^{-1-k-ip}
d^2 \mathbf{n}_z.
\nn
\eea
This integral can now be computed directly, the result being
\bea
\overline {\la w,\xi \ra}^{2k}   \repiso \towerbot \phi(w)
&=&  (-1)^{2k} \sqrt{  \frac{d_k }{\pi}} \frac{\sqrt{k^2+p^2}}{k-ip} |\la w,\xi \ra|^{4k}.
\nn
\eea
We cancel $\overline {\la w,\xi \ra}^{2k}$ from both sides
and use the homogeneity to evaluate the function for $|w| \neq 1$, since
\be
\phi(w) =   \phi\left(  \frac{w}{|w|} \right) |w|^{-2+2p}  = \phi\left(  \frac{w}{|w|}   \right) \la w,w\ra^{-1-ip}.
\ee
Which leaves us with
\bea
\repiso \towerbot \phi(w)
&=&  (-1)^{2k} \sqrt{  \frac{d_k }{\pi}} \frac{\sqrt{k^2+p^2}}{k-ip}   \la w,w\ra^{-1-ip-k}          \la w,\xi \ra^{2k}.
\nn
\eea
Taking the complex conjugate then gives equation \ref{Jformula}.

The action of $\bigJ^2$ gives
\bea
\label{Jsquaredformula}
\bigJ^2 \towerbot\phi(z)
&=& \frac{k^2+p^2}{(k+ip) (k-ip)   } \sqrt{  \frac{d_k }{\pi}}        \la z,z\ra^{-1+ip-k}[z,J^2 \xi]^{2k}
\nn \\
&=&
(-1)^{2k} \towerbot\phi(z)
\eea

\paragraph{Boundary data}
In the context of the four-simplex, the coherent state in tetrahedron $a$ along the dual edge leading to tetrahedron $b$ is denoted $\phi_{ab}(z)= \sqrt{  d_{k_{ab}} \pi^{-1}   } [z,\xi_{ab}]^{2k_{ab}}$.
The state for the tetrahedron is then obtained by tensoring together four coherent states, one for each triangle, and then averaging over $\SU(2)$ to obtain the $\SU(2)$-invariant part.
\begin{equation}\label{coherenttet}\hat\iota_a=\int_{\SU(2)}\dd g\;\bigotimes_{b\colon b\ne a} g \phi_{ab}.\end{equation}

This intertwiner is determined, up to phase, by the spins $k_{ab}$ and unit vectors $\mathbf n_{ab}=\mathbf n_{\xi_{ab}}$ for one fixed $a$ and the four values of $b$ satisfying $b\ne a$. This classical data is called {\em boundary data}, while the intertwiners themselves are called {\em boundary states}.

{\em Non-degenerate boundary data} is defined to be boundary data which is such that, for each $a$, the four vectors $\mathbf{n}_{ab}$, for $b \neq a$, span a three-dimensional space. Throughout this paper, we will assume that the boundary data is non-degenerate, which means that we will not cover a few extremal cases.

\paragraph{Propagators}
Applying the formula for the amplitude to these boundary states gives
\be\label{amplitude}
f_4 = (-1)^{\chi} \int_{\SL(2,\C)^5} \prod_{a} dX_a \, \delta(X_5) \prod_{a < b} P_{ab}
\ee
with $\chi$ determined by the graphical calculus, and the propagator defined as
$$    P_{ab}=\beta(\bar{X}_a\towerbot\phi_{ab},\bar{X}_b\towerbot\phi_{ba}).$$
In this formula, the integration over $\SU(2)$ in \eqref{coherenttet} can be ignored because it is subsumed in the integrals over the Lorentz group.

The propagator can be expressed as a double integral using \eqref{bilinear}. However, \eqref{Jformula} removes one of the integrals, leading to a more compact formula
\be
P_{ab}=c_{ab} d_{k_{ab}} \int_{\CP^1}\la X_a^{\dagger}z,X_a^{\dagger}z\ra^{-1-ip_{ab}-k_{ab}}\la X_a^{\dagger}z,\xi_{ab}\ra^{2k_{ab}}
\la X_b^{\dagger}z,X_b^{\dagger} z\ra^{-1+ip_{ab}-k_{ab}}\la J \xi_{ba},X_b^{\dagger}z\ra^{2k_{ab}} \;\Omega_z \nn
\ee
with $c_{ab}=\frac{ \sqrt{k_{ab}^2+p_{ab}^2}}{\pi   ( k_{ab}-ip_{ab})}$.  Note that this reduces to the formula in \cite{Barrett:1999qw} for $k_{ab}=0$.

The formula for the amplitude can be derived in different ways. One approach is via canonical bases, in the next section. Another approach is to generalise the construction of \cite{Barrett:1999qw} directly by replacing the functions on the unit hyperboloid by sections of tensor powers of the spinor bundle over the unit hyperboloid. The spin network vertex now requires a tensor at each point of the hyperboloid to contract together four sections to form a scalar which can be integrated over the hyperboloid. This tensor is an $\SU(2)$ tensor, since this is the group for the spin space at each point on the hyperboloid. The tensor is of course given by the $\SU(2)$ intertwiner $\hat\iota$.

\subsection{Canonical bases}

In this subsection we would like to establish a dictionary with the construction presented in \cite{Engle:2007wy}. There, the canonical bases for $\SL(2,\C)$ were heavily used and the language was more algebraic, in contrast to the more geometrical setting developed in this paper. The definition of the model used here is slightly different than the one in \cite{Engle:2007wy}.

Following \cite{GGV,naimark,ruhl}, we use the canonical basis $f^j_q(z)^{(k,p)}$ for the representation space $\rep_{(k,p)}$. For the moment we keep $(k,p)$ arbitrary, in particular, $k$ may be negative. As homogeneous functions, they can be given an explicit expression using hypergeometrical functions. However it will be more useful to use homogeneity to scale the argument such that it is normalized: $\la z , z\ra=1$. Given a normalized spinor $\xi$, construct the $\SU(2)$ matrix:

\be
g(\xi)=\left(\begin{array}{cc} \xi_0 & -\bar{\xi}_1 \\
                           \xi_1 & \bar{\xi}_0  \end{array}\right). \nonumber
\ee

The canonical basis when restricted to normalized spinors is identified with $\SU(2)$ representation matrices:
\be
f^j_q(\xi)^{(k,p)}= \sqrt{\frac{d_j}{\pi}}\, D^j_{q \, k}(g(\xi)). \nonumber
\ee
To evaluate $f^j_q(z)^{(k,p)}$ on non-normalized spinors we use homogeneity and the identification with $\SU(2)$ representation matrices given by the last formula. We then have:
\be
f^j_q(z)^{(k,p)}= \sqrt{\frac{d_j}{\pi}}\, \la z, z\ra^{ip-1-j} \; D^j_{q\, k}(g(z)),
\ee
where we have extended the action $g$ defined above to non-normalized spinors, the result of which lies in $GL(2,\C)$. One can check that this function has the correct homogeneity. The representation matrices are given explicitly by:
\bea
D^{j}_{q\, k}(g(z))&=&\left[\frac{(j+q)!(j-q)!}{(j+k)!(j-k)!}\right]^{\half}\; \sum_n \left(\begin{array}{c}j+k \\ n\end{array}\right)
\left(\begin{array}{c}j-k \\ j+q-n \end{array}\right) \times \nonumber \\ &\times&\left(z_0\right)^{n}\left(-\bar{z}_1\right)^{j+q-n}\left(z_1\right)^{j+k-n}\left(\bar{z}_0\right)^{n-q-k}. \label{Dmatrix}
\eea
In the expression above, the sum over $n$ is such that the binomial coefficients do not vanish. Coherent states $C^j_{\xi}(z)^{(k,p)}$ are defined by \footnote{In this section, for clarity, we use a different notation for the action of an $\SL(2,\C)$ group element on homogeneous functions of degree $(k,p)$: $\left(Xf\right)(z)=:T_{X}^{(k,p)}f \, (z)$. This notation leaves explicit the space on which the group element $X$ is being represented.}:
\be\label{cbcohst}
C^j_{\xi}(z)^{(k,p)}:=T_{g(\xi)}^{(k,p)}f^j_{j}(z)^{(k,p)}=f^j_{j}(g(\xi)^T\, z)^{(k,p)}=f^j_{j}\left(\la\bar{z},\xi\ra,\left[\bar{\xi}, z\right]\right)^{(k,p)},
\ee
where the notation $f^j_{j}(z) = f^j_{j}(z_1,z_2)$ has been used in the last step. Let us now see how the unitary isomorphism $\repiso$ acts on the coherent states defined above. Because $\repiso$ intertwines the representations $(k,p)$ and $(-k,-p)$ we have that:
\bea
\repiso C^j_{\xi}(z)^{(k,p)} &=& \repiso T_{g(\xi)}^{(k,p)}f^j_{j}(z)^{(k,p)} = T_{g(\xi)}^{(-k,-p)} \repiso f^j_{j}(z)^{(k,p)} = \nonumber \\
&& = \tilde{c} \, T_{g(\xi)}^{(-k,-p)}f^j_{j}(z)^{(-k,-p)}= \tilde{c} \, C^j_{\xi}(z)^{(-k,-p)}, \nonumber
\eea
where we used in the second to last step that the action of $\repiso$ commutes with the generators of $\SL(2,\C)$ and, in particular, maps elements of the basis in $\rep_{(k,p)}$ to elements of the basis in $\rep_{(-k,-p)}$, up to a constant $\tilde{c}$ that may depend on $j$ and $(k,p)$\footnote{Note that the normalization of $\repiso$ is arbitrary. This constant will be chosen to agree with the results of the last section.}.

We now restrict to the case $j=k$ with $k$ positive. In this case the sum in \eqref{Dmatrix} collapses to a single term. Then using \eqref{cbcohst} we have that
\be\label{cbcohstk}
C^k_\xi(z)^{(k,p)}= \sqrt{\frac{d_k}{\pi}}\, \la z, z\ra^{ip-1-k}\; \la\bar{z} ,\xi \ra^{2k}
\ee
and
\be
C^k_\xi(z)^{(-k,-p)}= \sqrt{\frac{d_k}{\pi}}\, \la z, z\ra^{-ip-1-k}\; D^k_{k\, -k}(g(\xi)^T\, z)= \sqrt{\frac{d_k}{\pi}}\, \la z, z\ra^{-ip-1-k}\; \left[\bar{z},\xi\right]^{2k}.
\ee
The action of $\repiso$ on a coherent state is then given explicitly by:
\be
\repiso \sqrt{\frac{d_k}{\pi}}\, \la z, z\ra^{ip-1-k}\; \la\bar{z} ,\xi \ra^{2k} = \tilde{c} \sqrt{\frac{d_k}{\pi}}\, \la z, z\ra^{-ip-1-k}\; \left[\bar{z},\xi\right]^{2k}. \nonumber
\ee
These formulas are to be compared with the ones obtained in the last section from explicit calculations. The coherent state \eqref{cbcohstk} is identified with $\towerbot\phi (z)$ by the change of variables:
\be
z = (z_0,z_1) \mapsto (-z_1,z_0) = J\bar{z}. \nonumber
\ee

The Lorentzian four-simplex amplitude defining the spin foam model is now defined. In the second part of this paper, we study the asymptotic properties of the amplitude $f_4$. More precisely, we study the regime in which the representation labels $(k,p)$ are large using extended stationary phase methods \cite{Hormander}.

\section{Symmetries and critical points}

Each propagator contains an internal variable, $z$, which is integrated over. Where it is necessary to distinguish these variables on the different propagators, the notation $z_{ab}$ will be used for this variable, for each $a<b$.
In the following, the combinations
 $$\z_{ab} = X_a^{\dagger}z_{ab} \quad\text{ and }\quad \z_{ba} = X_b^{\dagger}z_{ab}$$
for each $a<b$ occur frequently; this notation will be used as a shorthand.

Using this notation, the propagator can be written as
\be
P_{ab} = c_{ab} d_{k_{ab}} \int_{\CP^1} \Omega_{ab} \left( \frac{\la \z_{ba}, \z_{ba} \ra}{\la \z_{ab}, \z_{ab} \ra} \right)^{i p_{ab}}
\left( \frac{\la \z_{ab}, \xi_{ab} \ra \la J \xi_{ba}, \z_{ba} \ra}{\la \z_{ab}, \z_{ab} \ra^{1/2} \la \z_{ba}, \z_{ba} \ra^{1/2}} \right)^{2k_{ab}}, \nn
\ee
where
$$
\Omega_{ab} = \frac{\Omega}{\la \z_{ab}, \z_{ab} \ra \la \z_{ba}, \z_{ba} \ra},$$
which is a measure on $\CP^1$.
Therefore, the four-simplex amplitude can be re-expressed in a form amenable to stationary phase as follows
\be
f_4 = (-1)^{\chi} \int_{(\SL(2,\C))^5} \delta(X_5) \,\prod_{a} dX_a \,  \int_{(\CP^1)^{10}} \; e^S\;\prod_{a<b} c_{ab} d_{k_{ab}} \, \Omega_{ab} . \nn
\ee
The action $S$ for the stationary problem is given by
\be
\label{staction}
S[X,z] = \sum_{a<b} k_{ab} \log \frac{  \la \z_{ab}, \xi_{ab} \ra^2 \la J \xi_{ba}, \z_{ba} \ra ^2}{\la \z_{ab}, \z_{ab} \ra \la \z_{ba}, \z_{ba} \ra} + i  p_{ab} \log \frac{\la \z_{ba}, \z_{ba} \ra}{\la \z_{ab}, \z_{ab} \ra} .
\ee
The first term is complex, defined mod $2\pi i$,  and the second term purely imaginery.

\subsection{Symmetries of the action}\label{symmetriesaction}

It is straightforward to see that the amplitude \eqref{amplitude}, and also the action \eqref{staction} (modulo $2\pi i$), admit three types of symmetry.

\begin{itemize}
\item {\em Lorentz.} A global transformation $X_a \rightarrow Y X_a$, $z_{ab} \rightarrow (Y^{\dagger})^{-1} z_{ab}$, for $Y$ in
$\SL(2,\C)$, acting on all the variables simultaneously.
\item {\em Spin lift.} At each vertex $a$, there is a transformation which takes $X_a \rightarrow - X_a$ and leaves all other variables fixed (so $X_a \rightarrow  X_b$ for $b\ne a$).
\item {\em Rescaling.} At each triangle $a<b$, there is a transformation which takes $z_{ab} \rightarrow \kappa z_{ab}$ for $0\ne\kappa\in \mathbb{C}$ and leaves all other variables fixed.

\end{itemize}

Note however that the Lorentz symmetry does not affect the asymptotic problem because the amplitude is gauge-fixed such that $X_5 = 1$. The spin lift symmetry then only acts at the vertices $a=1,2,3,4$.

\subsection{Critical points}

If the representation labels $(k_{ab},p_{ab})$ assigned to the triangles of the $4$-simplex are simultaneously rescaled by a constant parameter $(k_{ab},p_{ab}) \rightarrow (\lambda k_{ab}, \lambda p_{ab})$, in the regime where $\lambda \rightarrow \infty$ the amplitude $f_4$ is dominated by the critical points of the complex action $S$, that is, the stationary points of $S$ for which $\mathrm{Re} \, S$ is a maximum. It is assumed from now on that $(k,p)\ne0$.

\subsubsection{Condition on the real part of the action}

The real part of the action
\be
\mathrm{Re} \, S = \sum_{a<b} k_{ab} \log \frac{ |\la \z_{ab}, \xi_{ab} \ra |^2 |\la J \xi_{ba}, \z_{ba} \ra |^2}{\la \z_{ab}, \z_{ab} \ra \la \z_{ba}, \z_{ba} \ra}, \nn
\ee
satisfies $\mathrm{Re} \, S\le0$ and is hence a maximum where it
vanishes. It vanishes if and only if, on each triangle $ab$, $a < b$, the following condition holds
\be
\label{real}
\frac{ \la \z_{ab}, \xi_{ab} \ra \la \xi_{ab}, \z_{ab} \ra \la J \xi_{ba}, \z_{ba} \ra \la \z_{ba}, J \xi_{ba} \ra }{\la \z_{ab}, \z_{ab} \ra \la \z_{ba}, \z_{ba} \ra} = 1.
\ee
This equation admits solutions if the coherent states $\xi_{ab}$ and $J \xi_{ba}$ are proportional to $\z_{ab}$ and $\z_{ba}$ respectively. Considering the fact that the coherent states are normalised, the most general solution to the above equation can be written
\be
\label{realsol}
\xi_{ab} = \frac{e^{i \phi_{ab}}}{\parallel \z_{ab} \parallel} X_a^{\dagger} \, z_{ab}, \;\;\;\; \mbox{and} \;\;\;\; J \xi_{ba} = \frac{e^{i \phi_{ba}}}{\parallel \z_{ba} \parallel} X_b^{\dagger} \, z_{ab},
\ee
where $\parallel \z_{ab} \parallel$ is the norm of $\z_{ab}$ induced by the Hermitian inner product, and $\phi_{ab}$ and $\phi_{ba}$ are phases. Eliminating $z_{ab}$, and introducing the notation $\theta_{ab} = \phi_{ab} - \phi_{ba}$, we obtain the first of the equations for a critical point
\be
\label{crit1}
(X_a^{\dagger})^{-1} \, \xi_{ab} = \frac{\parallel \z_{ba} \parallel}{\parallel \z_{ab} \parallel} e^{i \theta_{ab}} (X_b^{\dagger})^{-1} J \, \xi_{ba},
\ee
for each $a<b$. We now turn towards the variational problem of the action.

\subsubsection{Stationary and critical points}\label{scp}

We now compute the critical points of the action by evaluating the first derivative of the action of the configurations satisfying the condition \eqref{real}.
The action \eqref{staction} is a function of the $\SL(2,\C)$ group variables $X$ and of the spinors $z$. We start by considering stationarity with respect to the spinor variables.

There is a spinor $z_{ab}$ for each triangle $ab$, $a < b$, and the variation of the action with respect to these complex variables  gives two spinor equations for each triangle. For the triangle $ab$, the variation with respect to the corresponding $z$ variable leads to the following (co-)spinor equation
\bea
\delta_{z} S &=& i p_{ab} \left( \frac{1}{\la \z_{ba}, \z_{ba} \ra} (X_b \z_{ba})^{\dagger} - \frac{1}{\la \z_{ab}, \z_{ab} \ra} (X_a \z_{ab})^{\dagger} \right) \nn \\
&& + k_{ab} \left( \frac{2}{\la J \xi_{ba}, \z_{ba} \ra} (X_b J \xi_{ba})^{\dagger} - \frac{1}{\la \z_{ab}, \z_{ab} \ra} (X_a \z_{ab})^{\dagger}  - \frac{1}{\la \z_{ba}, \z_{ba} \ra} (X_b \z_{ba})^{\dagger} \right), \nn
\eea
while the variation with respect to $\bar{z}$ yields the spinor equation displayed below
\bea
\delta_{\bar{z}} S &=& i p_{ab} \left( \frac{1}{\la \z_{ba}, \z_{ba} \ra} X_b \z_{ba} - \frac{1}{\la \z_{ab}, \z_{ab} \ra} X_a \z_{ab} \right)
\nn \\ && + k_{ab} \left( \frac{2}{\la \z_{ab}, \xi_{ab} \ra} X_a \xi_{ab} - \frac{1}{\la \z_{ab}, \z_{ab} \ra} X_a \z_{ab}  - \frac{1}{\la \z_{ba}, \z_{ba} \ra} X_b \z_{ba} \right). \nn
\eea
Evaluating the above variations on the motion \eqref{realsol} and equating them to zero leads to the following two equations
\be
(X_a \, \xi_{ab})^{\dagger} = \frac{\parallel \z_{ab} \parallel}{\parallel \z_{ba} \parallel} e^{- i \theta_{ab}} (X_b \, J \, \xi_{ba})^{\dagger}, \;\;\;\; \mbox{and} \;\;\;\; X_a \, \xi_{ab} = \frac{\parallel \z_{ab} \parallel}{\parallel \z_{ba} \parallel} e^{i \theta_{ab}} X_b \, J \, \xi_{ba}, \nn
\ee
using the assumption that   $(k_{ab},p_{ab})\ne(0,0)$.
The two equations above are related by Hermitian conjugation and there is therefore only one relevant equation extracted from the stationarity of the spinor variables. Thus, our second critical equation is the following
\be
\label{crit2}
X_a \, \xi_{ab} = \frac{\parallel \z_{ab} \parallel}{\parallel \z_{ba} \parallel} e^{i \theta_{ab}} X_b \, J \, \xi_{ba}.
\ee

Finally, we consider stationarity with respect to the group variables. The right variation of an arbitrary $\SL(2,\C)$ element $X$ and its Hermitian conjugate are given by
\be
\delta X = X  L, \;\;\;\; \mbox{and} \;\;\;\; \delta X^{\dagger} = L^{\dagger}  X^{\dagger},
\ee
where $L$ is an arbitrary element of the real Lie algebra $\sl(2,\C)_{\R}$.

Because $\sl(2,\C) = \su(2)^{\C}$, there exists a vector space isomorphism $\sl(2,\C)_{\R} \cong \su(2) \oplus i \su(2)$. Using this isomorphism, we can decompose $L$ into a rotational and boost part $L = \alpha^i J_i + \beta^i K_i$, with $\alpha^i, \beta^i$ in $\R$ for all $i=1,2,3$. Throughout this paper, we will use the convention $\mathbf{J} = \frac{i}{2} {\pmb \sigma}$ and $\mathbf{K} = \frac{1}{2} {\pmb \sigma}$ for the spinor representation of the rotation and boost generators respectively. See Appendix B for further details on our conventions.

The variation of the action with respect to the group variable\footnote{The variation with respect to the variable $X_a$ performed here corresponds to a vertex which is the source of all its edges.  The variation for a vertex which
is the target of some, or all, of its edges will require varying with respect to $X_b$.
This proceeds similarly to the above and leads to the same stationary
point equations.} $X_a$, $a=1,...,4$, yields
\bea
\delta_{X_a} S &=& -\sum_{b : b \neq a} \left[ i p_{ab} \left( \frac{\la \z_{ab}, L \, \z_{ab} \ra}{\la \z_{ab}, \z_{ab} \ra} + \frac{\la \z_{ab}, L^{\dagger} \, \z_{ab} \ra}{\la \z_{ab}, \z_{ab} \ra} \right) \right. \nn \\&& \left. + k_{ab} \left( \frac{\la \z_{ab}, L \, \z_{ab} \ra}{\la \z_{ab}, \z_{ab} \ra} + \frac{\la \z_{ab}, L^{\dagger} \, \z_{ab} \ra}{\la \z_{ab}, \z_{ab} \ra} - 2 \frac{\la \z_{ab}, L \, \xi_{ab} \ra}{\la \z_{ab}, \xi_{ab} \ra} \right) \right]. \nn
\eea
Now, evaluating this first derivative on the points satisfying the condition on the real part of the action \eqref{realsol} and equating the result to zero leads to
\bea
\sum_{b : b \neq a} i p_{ab} \left( \la \xi_{ab}, L \, \xi_{ab} \ra + \la \xi_{ab}, L^{\dagger} \, \xi_{ab} \ra \right) + k_{ab} \left( - \la \xi_{ab}, L \, \xi_{ab} \ra + \la \xi_{ab}, L^{\dagger} \, \xi_{ab} \ra \right) = 0. \nn
\eea
Now we use the fact that the spinors $\xi_{ab}$ determine $\SU(2)$ coherent states, that is,
\be
\la \xi_{ab}, \mathbf{J} \, \xi_{ab} \ra = \frac{i}{2} \, \mathbf{n}_{ab}, \;\;\;\; \mbox{and} \;\;\;\; \la \xi_{ab}, \mathbf{K} \, \xi_{ab} \ra = \frac{1}{2} \, \mathbf{n}_{ab},
\ee
where $\mathbf{n}_{ab}\in \R^3$ is the unit vector corresponding to the coherent state $\xi_{ab}$. This leads immediately to
the following two equations
\be
\sum_{b : b \neq a} p_{ab} \mathbf{n}_{ab} = 0, \;\;\;\; \mbox{and} \;\;\;\;  \sum_{b : b \neq a} k_{ab} \mathbf{n}_{ab} = 0, \nn
\ee
because $\alpha^i$ and $\beta^i$ are arbitrary.
Since we are considering non-degenerate boundary data, these two equations can only be satisfied if there is a restriction on the representation labels.  We have that $p_{ab} = \gamma_a k_{ab}$ for
some arbitrary constant $\gamma_a$ at the a-th tetrahedron. However,
since the equations hold for each tetrahedron, $\gamma_a = \gamma_b =
\gamma$ and  the representations are related by a global parameter $\gamma$, i.e.,
\begin{equation}p_{ab} = \gamma k_{ab}.\label{Immirzi}\end{equation}

This provides further evidence for the
simplicity constraints given in \cite{Engle:2007wy} as we have shown
that the action does not admit any stationary points unless this
condition is satisfied \footnote{Note that this conclusion would no longer be valid if we were to consider degenerate boundary data. We thank the referee for pointing out this subtlety.}. Therefore, the two equations collapse to a single stationary point equation
\be
\label{crit3}
\sum_{b:b \neq a} k_{ab} {\bf n}_{ab} = 0.
\ee

Boundary data $\{k_{ab},\mathbf n_{ab}\}$ on a tetrahedron satisfying this equation determines a metric geometry for the tetrahedron and an orientation. This is the unique Euclidean geometry for the tetrahedron which has the $\mathbf n_{ab}$ as the outward-pointing normals and $k_{ab}$ the areas of the faces. Boundary data for a 3-manifold is said to be {\em Regge-like} if these geometries are non-degenerate and glue up to give a consistent Regge metric and an orientation for the whole 3-manifold \cite{Barrett:2009gg}.

To summarise, we have obtained four critical point equations given by expressions \eqref{crit1}, \eqref{crit2}, \eqref{Immirzi} and \eqref{crit3}. Solutions to these equations dominate the asymptotic formula for the Lorentzian $4$-simplex amplitude.

\section{Geometrical interpretation}

In this section, we show how geometrical structures emerge from the critical point equations. For the rest of the paper it is assumed that there is a fixed constant $\gamma$ such that $p_{ab} = \gamma k_{ab}$. Thus the four-simplex amplitude is determined by the boundary data and $\gamma$ alone.

\subsection{Bivectors}

Let $\Lambda^2(\R^{3,1})$ be the space of Lorentzian bivectors. A pair of vectors $N,M \in \R^{3,1}$ determines a simple bivector $N \wedge M$ which can be considered as the antisymmetric tensor
$$
N\wedge M=N\otimes M-M\otimes N.
$$
The above equation fixes our conventions for the wedge product of two vectors.
The norm $|B|$ of a bivector $B$ in $\Lambda^2(\R^{3,1})$ is defined by
$$
|B|^2 = \frac{1}{2} B^{IJ} B_{IJ},
$$
where $I,J,K=0,...,3$ label the components of the antisymmetric tensor, and indices are raised and lowered with the standard Minkowski metric $\eta = (-,+,+,+)$ on $\R^{3,1}$. A bivector is said to be space-like (resp. time-like) if $|B|^2 >0$ (resp. $|B|^2 <0$).
We will use the fact that the space $\Lambda^2(\R^{3,1})$ can be identified as a vector space with the Lie algebra $\so(3,1)$ of the Lorentz group using the isomorphism $\varsigma: \Lambda^2(\R^{3,1}) \rightarrow \so(3,1)$, $B \mapsto Id \otimes \eta \,(B)$, with the metric regarded as a map $\eta : \R^{3,1} \rightarrow (\R^{3,1})^*$. Hence, if $B$ is viewed as an anti-symmetric four-by-four matrix, the identification with a Lorentz algebra element yields
\be
\label{identification}
\left[ \begin{array}{cccc} 0 & b_1 & b_2 & b_3 \\ - b_1 & 0 & r_1 & r_2 \\ -b_2 & - r_1 & 0 & r_3 \\ -b_3 & - r_2 & - r_3 & 0 \end{array} \right] \mapsto \left[ \begin{array}{cccc} 0 & b_1 & b_2 & b_3 \\ b_1 & 0 & r_1 & r_2 \\ b_2 & - r_1 & 0 & r_3 \\ b_3 & - r_2 & - r_3 & 0 \end{array} \right].
\ee
The above properties essentially summarise the differences between Lorentzian and Euclidean bivectors. In particular, the bivector geometry theorem \cite{barrett-1998-39} for an Euclidean $4$-simplex holds also in a Minkowski space version, with suitable adjustments for the Minkowski metric. Let $\sigma$ be a 4-simplex in $\R^{3,1}$ with all faces spacelike, and let $N_a$ be the outward normal to the $a$-th face.
The bivector of the $ab$-th triangle of a 4-simplex in $\R^{3,1}$ is defined to be
\be\label{bivectorofsimplex} B_{ab}(\sigma)=k_{ab}\frac{*N_a\wedge N_b}{|*N_a\wedge N_b|},\ee
where $k_{ab}$ is the area of the triangle. The Lorentzian bivector geometry gives the constraints on the triangle bivectors $B_{ab}(\sigma)$ of the $4$-simplex $\sigma$. Conversely, it also gives the constraints for an arbitrary set of ten bivectors $B_{ab}$ to correspond to the triangle bivectors $B_{ab}(\sigma)$ of a Lorentzian geometric $4$-simplex $\sigma$. Using this Lorentzian bivector geometry theorem, we can provide a geometrical interpretation of the critical points as follows.

The geometry of the critical points is based on the identification between spinors and null vectors. Let $\Gamma : \R^{3,1} \rightarrow \mathbb{H}$; $x \mapsto \Gamma(x) = x^{0} 1 \!\! 1 + x^i \sigma_{i}$ be the vector space isomorphism between Minkowski space $\R^{3,1}$ and the space of two-by-two Hermitian matrices $\mathbb{H}$.
Through this isomorphism, the action of a Lorentz group element $\hat{X}$ on $\R^{3,1}$ lifts to the action of an $\SL(2,\C)$ element $X$ on $\mathbb{H}$ as follows $\Gamma (\hat{X} x) = X \Gamma(x) X^{\dagger}$.
Using this isomorphism, we can map spinors to null vectors through the following procedure. Let
\be
\zeta : \C^2 \rightarrow \mathbb{H}_0^+, \;\;\;\; z \mapsto \zeta(z) = z \otimes z^{\dagger}, \label{spinor to matrix map}
\ee
be the standard map between spin space and the space of degenerate two-by-two Hermitian matrices with positive trace
$$
\mathbb{H}_0^+ = \lbrace h \in \mathbb{H} \mid \det h = 0, \;\; \mbox{and} \;\; \mathrm{Tr} \; h >0 \rbrace.
$$
Note that this non-linear map is obviously not injective and satisfies $\zeta(r e^{i \theta} z) = r^2 \zeta(z)$.
Finally, the space $\mathbb{H}_0^+$ can be identified via $\Gamma$ to the future pointing null cone $C^+$ in Minkowski space and the map $\iota = \Gamma^{-1} \circ  \zeta : \C^2 \rightarrow \R^{3,1}$ maps spinors to null vectors.

Following the above construction, we can therefore associate the null vector
\be
\iota(\xi) = \frac{1}{2} (1,\mathbf{n}), \nn
\ee
to the coherent state $\xi$ by using equation \eqref{null}.
In fact, we can associate a second null vector to the coherent state $\xi$ by using the antilinear structure $J$:
\be
\iota(J \xi) = \frac{1}{2} (1, - \mathbf{n}). \nn
\ee
The two spinors $\xi$ and $J \xi$ form a spin frame because $[\xi, J \xi] = \langle J \xi , J \xi \rangle = 1$.

From this spin frame, we can construct bivectors in the vector representation as follows. Define the following time-like and space-like vectors
\be
\mathcal{T} = (\iota(\xi) + \iota(J \xi)) = (1, {\pmb 0}), \;\;\;\; \mbox{and} \;\;\;\;
\mathcal{N} = (\iota(\xi) - \iota(J \xi)) = (0, \mathbf{n}). \nn
\ee
From these two vectors construct the space-like bivector
\be
\label{northpole}
b = * \mathcal{T} \wedge \mathcal{N} = - 2 * \iota(\xi) \wedge \iota(J \xi),
\ee
where the star $*$ is the Hodge operator acting on the space $\Lambda^2(\R^{3,1})$.
In the four-by-four matrix representation, $b$ is given explicitly by
\be
b = * \left[ \begin{array}{cccc} 0 & n^1 & n^2 & n^3 \\ - n^1 & 0 & 0 & 0 \\ -n^2 & 0 & 0 & 0\\ -n^3 & 0 & 0 & 0 \end{array} \right].
\ee
This bivector is space-like because $\mathcal{T} \wedge \mathcal{N}$ is time-like and the Hodge operator is an anti-involution.
It is also simple by construction and satisfies a cross-simplicity condition
$$
\mathcal{T}^I b^{JK} \eta_{IJ} = 0.
$$
Following this construction for every coherent state $\xi_{ab}$ consequently leads to a collection of bivectors living in the hyperplane $\mathcal{T}^{\bot}$, with $\mathcal{T}$ the reference point of the future hyperboloid $H_3^+$.

Our critical point equations carry a richer structure. They correspond to conditions \cite{barrett-1998-39} satisfied by the bivectors $B_{ab}(\sigma)$ of a geometric $4$-simplex in $\R^{3,1}$. Consider the set of ten space-like bivectors constructed
as follows:
\be
\label{bivectors}
B_{ab} = k_{ab} \, \hat{X}_a \otimes \hat{X}_a \, b_{ab},
\ee
where $\hat{X}_a$ is the $\SO(3,1)$ element corresponding to $\pm X_a$ in $\SL(2,\C)$. These bivectors
are of norm $k_{ab}$ and satisfy the following bivector geometry conditions.

They are \emph{simple} and \emph{cross-simple} by construction. The cross-simplicity occurs because they live in the hyperplane $F_a^{\bot}$, with the future pointing vector $F_a$ in $H_3^+$ being the image of the reference vector $\mathcal{T}$ under the action of $X_a$, that is,
$$
F^I_a B^{JK}_{ab} \eta_{IJ} = 0, \;\;\;\; \mbox{with} \;\;\;\; \Gamma(F_a) = X_a X_a^{\dagger}.
$$
Note that the matrix $\Gamma(F_a)$ lives in the space $\mathbb{H}_1^+$ of two-by-two Hermitian matrices with unit determinant and positive trace, which is isomorphic to $H_3^+$ as a manifold.
Furthermore, the constructed bivectors satisfy \emph{closure}
$$
\sum_{b:b \neq a} B_{ab} = 0,
$$
because of the closure conditions \eqref{Immirzi} and \eqref{crit3} satisfied by the critical points.

Another important property is to show that the constructed bivectors also satisfy an \emph{orientation} equation
\be\label{orientationequation}
B_{ab} = - B_{ba}.
\ee
This is shown by the following argument.
Firstly, we compute the action of the $J$ structure on $\SL(2,\C)$.
Writing an arbitrary $\SL(2,\C)$ element $X$ as $X = \exp \, \alpha^i J_i + \beta^i K_i$, with $\alpha^i, \beta^i$ real, it is immediate to see that
\be
\label{P}
J X J^{-1} = (X^{\dagger})^{-1}.
\ee
The restriction of the above equation to the unitary subgroup states that $J$ commutes with $\SU(2)$ as expected.
Using this action, one can show that the two critical point equations \eqref{crit1}, \eqref{crit2} lead to the following equations in the vector representation of the Lorentz group
\be
\hat{X}_a \, \iota (\xi_{ab}) = \frac{\parallel \z_{ab} \parallel^2}{\parallel \z_{ba} \parallel^2} \, \hat{X}_b \, \iota ( J \xi_{ba}) , \;\;\;\; \mbox{and} \;\;\;\;
\hat{X}_a \, \iota (J \xi_{ab}) = \frac{\parallel \z_{ba} \parallel^2}{\parallel \z_{ab} \parallel^2} \, \hat{X}_b \, \iota (\xi_{ba}). \nn
\ee
Wedging these two vector equations leads to the bivector equation
\be
\hat{X}_a \otimes \hat{X}_a \, \iota (\xi_{ab}) \wedge \iota (J \xi_{ab}) = - \hat{X}_b \otimes \hat{X}_b \, \iota (\xi_{ba}) \wedge \iota (J \xi_{ba}),
\ee
which proves that the orientation condition \eqref{orientationequation}.

Finally, the \emph{tetrahedron} condition, which states that each tetrahedron has a non-degenerate geometry, is also fulfilled if we choose non-degenerate boundary data.

Therefore, for non-degenerate boundary data, the constructed bivectors \eqref{bivectors} satisfy five of the six bivector geometry constraints of \cite{barrett-1998-39}. The one remaining condition is \emph{non-degeneracy} of the bivector geometry, which states that the bivectors span the 6-dimensional space of bivectors.  Lemma 3 of \cite{Barrett:2009gg}, which states that the bivector geometry is either  non-degenerate or contained in a 3-dimensional hyperplane, did not use the metric at all, and so applies unaltered to our case. We shall first discuss the case of a non-degenerate bivector geometry which corresponds to a 4-simplex with a Lorentzian geometry.

\subsection{Lorentzian 4-simplex from non-degenerate bivector geometry}

The reconstruction theorem of \cite{barrett-1998-39} is the same as that used in \cite{Barrett:2009gg} for a Euclidean metric. The only difference is that when comparing the bivectors of the reconstructed 4-simplex to the bivector geometry, the Lorentzian metric is used to evaluate areas.

We therefore again have the equivalence of solutions to \eqref{crit1} to geometric 4-simplices up to inversion. Inversion is the isometry of Minkowski space given by $x\to -x$.
In particular, for every solution that is non-degenerate, there exists a parameter $\mu$ which takes the value either $1$ or $-1$, and an inversion-related pair of Lorentzian 4-simplexes $\sigma$.  These are such that the bivectors (of either of the two simplexes) $B_{ab}(\sigma)$ satisfy
\be
B_{ab}(\sigma) = \mu B_{ab} = \mu\, k_{ab} \, \hat{X}_a \otimes \hat{X}_a \, b_{ab} \label{geom-bivector}.
\ee

A key subtlety in the geometric interpretation of our equations arises due to the fact that $\SLtwoC$ maps only to the connected component of $\SO(3,1)$ and takes future pointing vectors into future pointing vectors. This means that the inversion map is not in $\SL(2,\C)$, and so the analysis in the following sections is somewhat different from the Euclidean case analysed in \cite{Barrett:2009gg}.

\subsection{Vector geometry from degenerate bivector geometry}

If the solutions fall into the case of degenerate bivector geometry, this implies all $F_a$ are pointing in the same direction. As we have fixed $X_5 = 1$ this means we have $F_a = X_a \pole = \pole$, for all $a$. That is the $X_a$ are in the $\SU(2)$ subgroup that stabilizes $\pole$. As such the two distinct critical and stationary point equations leading to the bivector equation \eqref{crit1} reduce to the single equation:
\be\label{vg}
X_a \, \xi_{ab} = e^{i \theta_{ab}} X_b \, J \xi_{ba}
\ee
The solutions to these equations have been studied initially in \cite{Barrett:2009gg} and more completely in \cite{su2paper}. A solution determines a geometrical structure called a vector geometry. This is a set of vectors $\mathbf v_{ab}\in\R^3$ satisfying closure, $\sum_a \mathbf v_{ab}=0$, and orientation, $\mathbf v_{ab}=-\mathbf v_{ab}$. In this case, $\mathbf v_{ab}=k_{ab}X_a \mathbf n_{ab}$.

\subsection{Symmetries and classification of the solutions}

An important input for the asymptotic formula is the classification of the solutions to the critical points. To start with, we need to consider their symmetries.

\subsubsection{Symmetries induced by the symmetries of the action}

The symmetries of the action listed in section \ref{symmetriesaction} map critical points to critical points, except that, as mentioned in section \ref{symmetriesaction}, some of the symmetries are broken by the gauge-fixing that is used to define the non-compact integrals.

One difference to the Euclidean case of \cite{Barrett:2009gg} is that there are only half as many discrete symmetries here as there were for the Euclidean amplitude. This means we are no longer able to invert the normal vectors using a different choice of lift.  The bivectors $B_{ab}$ are also left invariant by this symmetry as can be seen from equation \eqref{northpole}.

\subsubsection{Parity}\label{parity}

An additional symmetry of the critical points that is not determined by a symmetry of the action is the parity operation, given by the inversion of the spatial coordinates of Minkowski space.
Using the $\SU(2)$ antilinear map $J$, one can construct a map acting on Minkowski vectors through their identification with two-by-two Hermitian matrices.
Most importantly, because the $J$ map commutes with $\SU(2)$, it necessarily anticommutes with Hermitian matrices which implies that
\be
J \Gamma(x) J^{-1} = x^{0} 1 \!\! 1 - x^i \sigma_{i} =  \Gamma(Px),
\ee
where $P$ is the  mapping $(x^0,\mathbf{x}) \mapsto (x^0, - \mathbf{x})$ on $\R^{3,1}$.
Now, we have seen that the $J$ map has a well defined action \eqref{P} on $\SL(2,\C)$.
The corresponding parity transformation $P$ is of key importance since it shows that given a solution to the critical point equations \eqref{crit1} and \eqref{crit2}, the transformation
\be\begin{aligned}
X_a &\rightarrow P(X_a) = J X_a J^{-1} \quad \quad &\forall a,\\
z_{ab} &\rightarrow  X_a X_a^\dag z_{ab} \ \ &\forall a<b
\end{aligned}\ee
leaves the critical point equations unchanged, because $\parallel \z_{ab} \parallel / \parallel \z_{ba} \parallel$ is mapped to $(\parallel \z_{ab} \parallel / \parallel \z_{ba} \parallel)^{-1}$, and is an involution. Therefore, $P$, together with the above transformation on $z$ is a symmetry of the critical points. It is not a symmetry of the action, but a prescription to construct a solution out of a solution.

An important feature of the action of $P$ is that it flips the orientation parameter $\mu$ to $-\mu$. The remainder of this section shows that this is the case.

Firstly, from the definition \eqref{bivectorofsimplex} of the bivectors of a 4-simplex, it follows that
$$ PB_{ab}(\sigma)=-B_{ab}(P\sigma).$$
This is because the the normal vectors $N_a$ transform as vectors under $P$, but $*P=-P*$. Another way of saying this is that the $N_a$ are determined by the metric only but $*$ requires an orientation.

Secondly, the bivector $b_{ab}$ has only space-space components, so is unaffected by the parity operation. This means that in the equation \eqref{geom-bivector} defining $\mu$,
$$P B_{ab}(\sigma)= \mu\, k_{ab} \, P(\hat{X}_a) \otimes P(\hat{X}_a) \, b_{ab} $$
and hence
$$\mu_\sigma=-\mu_{P\sigma}.$$

\subsubsection{Classification}\label{class-solutions}
 The solutions to the critical point equations are classified according to the different types of boundary data. In this classification, solutions which are related by the symmetries of the action, as described in section \ref{symmetriesaction}, are regarded as the same solution. For the Lorentzian 4-simplex, the results are described in the previous sections. The complete classification of the vector geometries for Regge-like boundary data is given in \cite{su2paper}.

\paragraph*{Regge-like boundary data.}
Given a Regge-like boundary geometry, a flat metric geometry for the 4-simplex is specified completely up to rigid motion. In particular the metric on the interior is uniquely fixed by knowing all the edge lengths, and this is fixed by the boundary data. One way to see this is to note that there is a linear isomorphism between the set of square edge lengths and set of interior metrics. Note that $l_{ab}^2 = v_{ab}^\mu v_{ab}^\nu g_{\mu \nu}$ for a 4-simplex with edge vectors $v$ and lengths $l$ is indeed linear and that choosing the edge vectors as basis vectors one can calculate their inner products using only the other edge lengths. But knowing the inner products and the lengths of the basis vectors is equivalent to knowing the metric. Now as the tetrahedra are all Euclidean and non-degenerate, the metric of the four-simplex must be of signature $(-,+,+,+)$, $(+,+,+,+)$ or $(0,+,+,+)$.

The solutions can therefore be classified according to the boundary data.
\begin{itemize}
\item {\em Lorentzian 4-simplex.} If the boundary data is that of a non-degenerate Lorentzian 4-simplex, then two distinct critical points exist, related by the parity transformation $P$ in section \ref{parity}.  Since the boundary data determines the metric of the 4-simplex $\sigma$, there are only four possibilities which are unrelated by the action of $\SL(2,\C)$, corresponding to the four connected components of the group $\OO(3,1)$. These are $\sigma$, its inversion partner $-\sigma$ and the parity-related $P\sigma$ and $-P\sigma$. The solutions correspond to inversion-related pairs, thus it is clear that the two solutions given by $(\sigma,-\sigma)$ and  $(P\sigma,-P\sigma)$ exhaust all the possibilities.

\item {\em 4d Euclidean 4-simplex.}
If the boundary data describes a  4-simplex in four-dimensional Euclidean space, then there will be exactly two critical points, $\{X^+_a\}$ and $\{X^-_a\}$, with all matrices in $\SU(2)$. These
can be  used to reconstruct a Euclidean 4-simplex $\sigma_E$, as in \cite{Barrett:2009gg}.
These critical points can also be used to construct the parity related 4-simplex $P\sigma_E$.

\item {\em 3d Euclidean 4-simplex. }
If the boundary data corresponds to a degenerate 4-simplex in $ \mathbb{R}^3$ then there will be a single $\SU(2)$ critical point.  This determines a vector geometry.  A second critical point cannot exist or we would be able to construct a non-degenerate Euclidean 4-simplex, which is not possible with this boundary data.

\end{itemize}

\paragraph*{Non Regge-like boundary data.}
If the boundary data is not Regge-like then the remaining possibilities are to obtain exactly one critical point in $\SU(2)$ which determines a vector geometry, or no critical points at all.

\section{Asymptotic formula}

In this section, we provide the asymptotic formula for the four-simplex amplitude $f_4$. We start by defining a canonical choice of phase for the boundary coherent state. Then, we evaluate the action on the critical points for boundary data forming a non-degenerate Lorentzian $4$-simplex geometry, and finally give the asymptotic formula.

\subsection{Regge states}

For Regge-like boundary data, the canonical choice of phase for the quantum state is called the Regge state and is defined in \cite{Barrett:2009gg}. This is as follows. Let $\sigma\subset\R^4$ be a 4-simplex (without any particular metric geometry or orientation), and $\tau_a$ the $a$-th tetrahedron. Denote a triangle in this 4-simplex by $\Delta_{ab}=\tau_a\cap\tau_b$. The boundary data determines, for each tetrahedron, an  affine linear map
$$\phi_a\colon \tau_a\to \R^3.$$ 
This is the map which takes $\tau_a$ to the tetrahedron $t_a\subset\R^3$ which has $\mathbf n_{ab}$ as the outward normal of the $b$-th face (as in section \ref{scp}). In the following, only the linear parts of these affine maps play any role, and the translational part is often ignored. In other words, the same notation is used for the derivative of this map of tangent spaces $T\tau_a\to\R^3$.

The boundary data determines \emph{gluing maps} $\hat g_{ab}\in\SO(3)$ by the requirements
$$\hat{g}_{ab} \bn_{ab} = - \bn_{ba}$$ and $$\hat{g}_{ab} \phi_a\bigl(\Delta_{ab}\bigr) = \phi_b\bigl(\Delta_{ba}\bigr).$$

These rotations can be lifted to $g_{ab}\in\SU(2)$ using the procedure of picking spin frames on each tetrahedron, as described in \cite{Barrett:2009gg}. Using these gluing maps the phase convention defining the Regge state is
\be\label{Regge state}
g_{ab} \xi_{ab} = J \xi_{ba}.
\ee

\subsection{Dihedral angles and boosts}
In this section and also in section \ref{sectionreggeaction}, critical points $X_a$ that correspond to a Lorentzian or 3d Euclidean 4-simplex are considered. The boundary data is therefore Regge-like, and a Regge state is assumed.

The following definitions of the dihedral angles and the Regge action for a Lorentzian simplex are based on the discussion in \cite{Barrett:1993db}. For a 4-simplex in Minkowski space with all tetrahedra space-like, the dihedral angles are all boost parameters. The $a$-th tetrahedron has an outward-pointing timelike normal vector $N_a$, and the dihedral angle at the intersection of two tetrahedra is determined up to sign by
$$\cosh\Theta_{ab}=|N_a\cdot N_b|,$$
and can be viewed as a distance on the unit hyperboloid.

\begin{figure}
\begin{center}
\includegraphics[height=5cm,width=10cm]{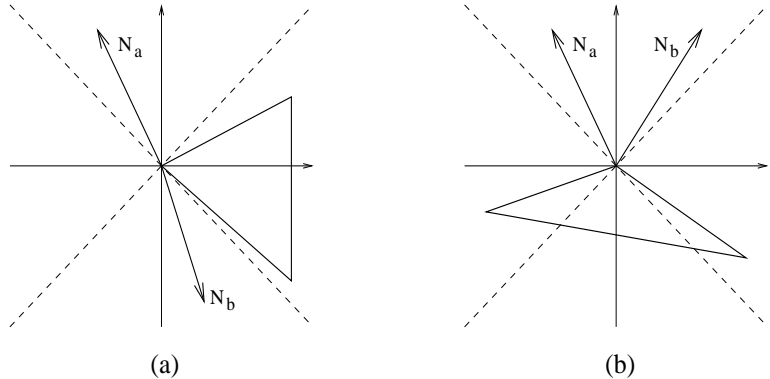}
\caption{Example of a thin wedge (a), and of a thick wedge (b).}\label{wedges}
\end{center}
\end{figure}

The sign of the dihedral angle is more delicate. One could define them all to be positive, but this would lead to additional signs in the formula for the Regge action. It is much better to take account of the nature of the triangle where the two tetrahedra meet. The tetrahedra come in two types: the outward normals are either future-pointing or past-pointing. The triangles are then classified into two types: {\em thin wedge}, where one of the incident tetrahedra is future and the other one past, and {\em thick wedge}, where both are either future or past (see figure \ref{wedges}).
The dihedral angle is defined to be positive for a thin wedge and negative for a thick wedge. It is worth noting that the type of the triangle is unchanged if $\sigma$ is replaced by $-\sigma$, so that the dihedral angle is an unambiguous property of a critical point.

The dihedral angles are related to the critical point data $X_a$ and the gluing maps. This proceeds via a modification of the formalism for Euclidean space. The reconstruction theorem determines an embedding of
the 4-simplex $\sigma$  in Minkowski space (this includes the case of 3d Euclidean geometry, in which $\sigma$ lies in a hyperplane). The Minkowski space has a standard orientation, so $\sigma$ and its boundary inherit an orientation from Minkowski space. This may or may not agree with the one induced by the boundary data using the maps $\phi_a$; this discrepancy is measured by the parameter $\mu$.

In this situation, the maps $\phi_a$ can be extended to elements $\Phi_a$ of $\SO(3,1)$ by mapping the outward unit normal $N_a$ to the vector $\pm \pole=\pm(1,0,0,0)$ and the tetrahedron $\tau_a$ to the orthogonal plane using $\phi_a$. The sign here is fixed by the requirement that $\Phi_a$ be orientation-preserving. As critical points only exist when the boundary data induces a coherent orientation around the 4-simplex (and are thus Regge-like), the sign is the same for all $\Phi_a$, that is, this extension maps all normals to the same vector $\pm \pole$ depending on whether the 4-simplex boundary agrees with the one induced by the embedding in Minkowski space or not.


Now define $\hat E_{ab}\in\SO(3,1)$ with the aid of the commutative diagram
$$
\xymatrix{\ar @{} [dr]  \R^{3,1} \ar[d]_{\hat g_{ab}} && \ar[ll]^{\Phi_a}  ~\R^{3,1} \ar[d]^{ \hat E_{ab}}   \\
\R^{3,1} && \ar[ll]_{\Phi_b}  ~\R^{3,1}  }
$$

This has the properties that $\hat E_{ab}(N_a)=N_b$ and all points of the triangle $\Delta_{ab}$ are fixed. In the thick wedge case this means that $\hat E_{ab}$ is a pure boost in the plane orthogonal to $\Delta_{ab}$, but in the case of a thin wedge $\hat E_{ab}$ reverses the time orientation, so is not an element of $\SO^+(3,1)$. However in this case, $-\hat E$ is now an element of $\SO^+(3,1)$. This acts as a pure boost in the plane orthogonal to $\Delta_{ab}$ composed with a rotation by $\pi$ in the plane of the triangle. (A rotation by $\pi$ in a two-dimensional plane acts as minus the identity.)

Therefore $\hat E_{ab}$ has a decomposition into a rotation $\hat R_{ab}$ and a pure boost $\hat D_{ab}$ according to
$$\hat E_{ab}=\begin{cases}\hat  D_{ab}\hat  R_{ab}\quad &\text{thick wedge}\\
-\hat D_{ab}\hat R_{ab} &\text{thin wedge}\end{cases}
$$
In the first case, $\hat R_{ab}$ is the identity and in the second case the rotation by angle $\pi$. The boost $\hat D_{ab}$ has the properties that $\hat D_{ab}(F_a)=F_b$ and all points of $\Delta_{ab}$ are fixed.

Now consider a stationary point $\hat X_a$, for $a=1,\ldots,5$. The reconstruction theorem shows that the Lorentz transformation $\hat X_a\in\SO^+(3,1)$ maps the bivectors of
$\tau_a$ to the bivectors of $t_a$ up to an overall sign. Thus it seems reasonable that $\hat X_a$ should map $\tau_a$ to plus or minus $t_a$. This is the case, and the overall sign is fixed by

\begin{lem}
Defining the signs $\epsilon_a=\pm1$ by $N_a=\epsilon_a F_a$, then
$$\Phi_a= -\mu\epsilon_a \hat X_a^{-1}.$$
\end{lem}

\emph{Proof.} Let $\sigma$ be the 4-simplex in Minkowski space with outward timelike unit normals $N_a$ for its tetrahedral faces. Then if $m_{ab}\in\R^{3,1}$ is the spacelike unit normal vector outward to the face $\tau_a\cap\tau_b$ of the tetrahedron $\tau_a$ and orthogonal to $N_a$, then it follows that $m_{ab}\cdot N_b<0$. This is because resolving $m_{ab}$ into components parallel and perpendicular to $N_b$, the component parallel to $N_b$ is a positive multiple of $N_b$ (and $N_b^2<0$).

Now from the reconstruction theorem,
$$\epsilon_a N_a\wedge\hat X_a(0,\mathbf n_{ab})=\mu \frac{N_a\wedge N_b}{|*N_a\wedge N_b|}$$
from which it follows that $\hat X_a(0,\mathbf n_{ab})=\pm m_{ab}$.
Dotting both sides with $N_a\wedge\hat X_a(0,\mathbf n_{ab})$ gives
$$-\epsilon_a=-\mu N_b\cdot \frac{\hat X_a(0,\mathbf n_{ab})}{|*N_a\wedge N_b|}$$
and so $\hat X_a(0,\mathbf n_{ab})= -\mu\epsilon_a m_{ab}$. Since a tetrahedron is determined by the outward normals of its faces, this implies the result.

Using the lemma, the commuting diagram can now be written in terms of the critical points $\hat X_a$, as
$$
{
\xymatrix{\ar @{} [dr] \R^{3,1} \ar[d]_{\hat g_{ab}} \ar[rr]^{ \hat X_a} && ~\R^{3,1} \ar[d]^{ \hat D_{ab} \hat R_{ab}   }   \\
\R^{3,1} \ar[rr]_{ \hat X_b} && ~\R^{3,1}  }  
}
$$
This commutes because the sign difference between $\hat E_{ab}$ and $\hat D_{ab} \hat R_{ab}$ is exactly $\epsilon_a\epsilon_b$.

The group elements $\hat D_{ab}$ and  $\hat R_{ab}$ can be lifted to $D_{ab}$ and  $R_{ab}$ respectively in $\SLtwoC$. The sign ambiguity of each of these needs to be fixed. The product of the two is defined by the commuting diagram
\begin{equation}\label{commuting}
\xymatrix{\ar @{} [dr] \R^{3,1} \ar[d]_{g_{ab}} \ar[rr]^{    X_a} && ~\R^{3,1} \ar[d]^{  D_{ab} R_{ab}  }   \\
\R^{3,1} \ar[rr]_{   X_b} && ~\R^{3,1}  }
\end{equation}
Since $D_{ab}$ is a lift of a boost, it has real eigenvalues. The ambiguity in the sign of $D_{ab}$ is then fixed by requiring it to have positive eigenvalues. Then the definition of both matrices is fixed uniquely. The eigenvalues of the matrix $R_{ab}\in\SU(2)$ will be either $1$ or $-1$ for a thick wedge, or $\pm i$ for a thin wedge. These signs are determined more precisely below.


\subsection{Regge action}\label{sectionreggeaction}

In this section, the action \eqref{staction} at a critical point is expressed in terms of the underlying geometry. On all critical points, the real part vanishes and, using \eqref{realsol}, the imaginary part is
\be
\label{im part of action}
S = i \sum_{a<b} p_{ab} \, \log \, \frac{\parallel \z_{ba} \parallel^2}{\parallel \z_{ab} \parallel^2} + 2 k_{ab} \, \theta_{ab}.
\ee
In the case where the critical points determine a non-degenerate $4$-simplex, the following discussion shows that the argument of the logarithm is related to the  dihedral angle $\Theta_{ab}$ at the $ab$ triangle \cite{Barrett:1993db}.

If $F_a$ and $F_b$ are the future pointing normals determining the two hyperplanes intersecting along the triangle $ab$, the corresponding dihedral angle $\Theta_{ab}$ obeys
\be
\label{dihedral}
\cosh  \Theta_{ab} =  - F_a \cdot F_b = \frac{1}{2} \mathrm{Tr} \; (\Gamma(F_a)^{-1} \Gamma(F_b)).
\ee
To see how this relates to the critical points, couple equations \eqref{crit1} and \eqref{crit2} and eliminate $J \xi_{ab}$. This leads to the following eigenvalue equation
$$
X_a^{-1} X_b X_b^{\dagger} (X_a^{\dagger})^{-1} \; \xi_{ab} = \frac{\parallel \z_{ba} \parallel^2}{\parallel \z_{ab} \parallel^2} \; \xi_{ab}.
$$
The matrix $X_a^{-1} X_b X_b^{\dagger} (X_a^{\dagger})^{-1}$ in this equation is Hermitian, so it follows that it has eigenvalues
$$
e^{r_{ab}} = \frac{\parallel \z_{ba} \parallel^2}{\parallel \z_{ab} \parallel^2}.
$$
and the inverse of this, $e^{-r_{ab}}$. Moreover, the eigenvectors are orthogonal. Therefore the trace of this matrix is the same as the trace in \eqref{dihedral}, and so
\be\label{dihedraluptosign}
|\Theta_{ab}| =  |r_{ab}|.
\ee
Therefore, the parameter $r_{ab}$ is the dihedral angle up to a sign.

To solve this sign ambiguity, it will prove useful to obtain an exponentiated form of this matrix. This is achieved by noting that since the spinors $\xi_{ab}$ are $\SU(2)$ coherent states, they satisfy
\be
(\mathbf{J} . \mathbf{n}) \, \xi = \frac{i}{2} \, \xi, \;\;\;\; \mbox{and} \;\;\;\; (\mathbf{K} . \mathbf{n}) \, \xi = \frac{1}{2} \, \xi. \nn
\ee
Hence, the Hermitian matrix $X_a^{-1} X_b X_b^{\dagger} (X_a^{\dagger})^{-1}$ can be written as a pure boost
\be
\label{littled}
X_a^{-1} X_b X_b^{\dagger} (X_a^{\dagger})^{-1} = e^{2 r_{ab} \, \mathbf{K} . \mathbf{n}_{ab}}. 
\ee

Using the above expression, we can now overcome the sign ambiguity of \eqref{dihedraluptosign}, by using the definition of the dihedral angle of a Lorentzian $4$-simplex $\sigma$ in terms of the parameter of the dihedral boost. Explicitly, the dihedral boost around the triangle $ab$ is obtained as the exponential of the Hodge dual of the (normalized) bivector $B_{ab}(\sigma)$, after the identification $\varsigma : \Lambda^2(\R^{3,1}) \rightarrow \so(3,1)$ defined in \eqref{identification}:
\be
\hat{D}_{ab} = \exp \left( \Theta_{ab} \, \varsigma (* \tilde{B}_{ab}(\sigma)) \right),
\ee
where $\tilde{B}_{ab} = B_{ab} / |B_{ab}|$ is the normalised bivector.

This formula is proved as follows. Since $B_{ab}(\sigma)$ is a simple spacelike bivector, then $\hat{D}_{ab}$ is a Lorentz transformation which stabilises a spacelike plane. The bivector in the exponent is
$$\Theta_{ab} \, * \tilde{B}_{ab}(\sigma)= -\Theta_{ab}\frac{N_a\wedge N_b}{|*N_a\wedge N_b|}
=|\Theta_{ab}|\frac{F_a\wedge F_b}{|*F_a\wedge F_b|},$$
using the sign convention in the definition of a dihedral angle.
This bivector acts in the plane spanned by $F_a$ and $F_b$ and the boost parameter has the right magnitude. It just remains to check that it maps $F_a$ to $F_b$, and not vice-versa. To first order in small $\Theta$, one has
$$\exp\left(|\Theta_{ab}|\frac{\varsigma(F_a\wedge F_b)}{|*F_a\wedge F_b|}\right)\;F_a \simeq F_a+\frac{|\Theta_{ab}|}{|\sinh\Theta_{ab}|}\left((F_b\cdot F_a)F_a- (F_a\cdot F_a) F_b\right)\simeq F_b.$$
This calculation uses the convention replacing wedge products with bivectors, and the fact that $F^2=-1$.

Now, the expression \eqref{geom-bivector} of the geometric bivectors in terms of the the boundary data leads to the following equality
\be
\varsigma(* \tilde{B}_{ab}(\sigma)) = - \mu \, \hat{X}_a \left[ \begin{array}{cccc} 0 & n_{ab}^1 & n_{ab}^2 & n_{ab}^3 \\ n_{ab}^1 & 0 & 0 & 0 \\ n_{ab}^2 & 0 & 0 & 0\\ n_{ab}^3 & 0 & 0 & 0 \end{array} \right] \hat{X}_a^{-1} = \mu \, \hat{X}_a \, \pi( \mathbf{n}_{ab} \cdot \mathbf{K}) \, \hat{X}_a^{-1},
\ee
where $\pi : \sl(2,\C)_{\R} \rightarrow \mathrm{End} (\R^{3,1})$ is the vector representation of the Lorentz algebra (see equation \eqref{vectorrep} in Appendix B).

This implies that the previously-defined lift of the dihedral boost to $\SL(2,\C)$ is given explicitly by
\be
D_{ab} =  X_a \, e^{\mu \Theta_{ab} \mathbf{K} . \mathbf{n}_{ab}} \, X_{a}^{-1}.
\ee
 In this formula, the boost generators $\mathbf K$ are  $2\times2$ Hermitian matrices so that this gives an element of $\SL(2,\C)$ with positive eigenvalues. 

Next, we use that a property of the dihedral boost just established,
$$
\Gamma(\hat{D}_{ab} \, F_{a}) = \Gamma(F_b) .
$$
This can be written
$$D_{ab}X_aX_a^\dagger D_{ab}^\dagger=X_bX_b^\dagger,$$
which implies
\be
X_a \, e^{2 \mu \Theta_{ab} \mathbf{K} . \mathbf{n}_{ab}} \, X_a^{\dagger} = X_b \, X_b^{\dagger}. \nn
\ee
Comparing the last equality and \eqref{littled} finally gives that
\be
e^{2 \mu \Theta_{ab} \mathbf{K} . \mathbf{n}_{ab}} = e^{2 r_{ab} \mathbf{K} . \mathbf{n}_{ab}},
\ee
which implies that $\mu \Theta_{ab} = r_{ab}$. This solves the sign ambiguity and identifies the log term in the action \eqref{im part of action}.

The remaining task is to identify $\theta_{ab}$ in terms of the geometry. According to \eqref{crit2}, \eqref{Regge state} and \eqref{commuting},
  \bea
X_a \, \xi_{ab} &=
& \frac{\parallel \z_{ab} \parallel}{\parallel \z_{ba} \parallel} e^{i \theta_{ab}} X_b \, J \, \xi_{ba},\nn\\
&=& e^{-\mu\Theta_{ab}/2} e^{i \theta_{ab}} X_b \, g_{ab} \, \xi_{ab},\nn\\
&=&  e^{-\mu\Theta_{ab}/2}  e^{i \theta_{ab}} D_{ab} R_{ab} \, X_a \, \xi_{ab},
\eea
In particular, it follows that $ e^{i \theta_{ab}}$ is one of the eigenvalues of $R_{ab}$. Define the angle 
$$\Pi_{ab}=\begin{cases}0\quad &\text{thick wedge}\\
\pi&\text{thin wedge}\end{cases}$$
Then with this notation,
\begin{equation}\label{Rsign} X_a^{-1}R_{ab}X_a=\pm e^{\Pi_{ab} \mathbf J. \mathbf n_{ab}}.\end{equation}
The sign in this formula is identified as follows. The Levi-Civita connection varies continuously as the geometry is deformed to a 3d configuration in which all $\Theta_{ab}$ are zero, i.e., the 4-simplex is squashed to lie in a 3d space-like hypersurface in Minkowski space. The lifts to the spin group can also be taken so that they vary continuously. In particular, this means that the signs in \eqref{Rsign} are the same as in the 3d configuration.
 This 3d configuration is common to the case of a Euclidean 4-simplex as analysed in  \cite{Barrett:2009gg}. In this case, the $\Pi_{ab}$ are the dihedral angles of the four-simplex considered as a degenerate 4-simplex in Euclidean geometry, and the results of \cite{Barrett:2009gg} imply that 
\begin{equation}\label{Rformula}X_a^{-1}R_{ab}X_a=\epsilon_a\epsilon_b e^{\Pi_{ab} \mathbf J. \mathbf n_{ab}}\end{equation}
for some $\epsilon_a=\pm1$. Moreover, by using the spin lift symmetry, the action is the same as for a critical point at which $\epsilon_a=1$ for all $a$.

Therefore, the action \eqref{im part of action} on the critical points corresponding to a non-degenerate $4$-simplex yields the following formula in terms of a Lorentzian $4$-simplex determined by the boundary data
\be
S = i \mu \gamma\sum_{a<b}  k_{ab} \Theta_{ab} - i \sum_{a<b} k_{ab}\Pi_{ab}.
\ee
The first term in the formula is $i\mu$ times the Regge action for a Lorentzian 4-simplex with space-like tetrahedra and triangle areas $\gamma  k_{ab}$. Since the sum of the $k_{ab}$ at each tetrahedron is an integer,  the second term contributes a factor $i\pi$ times an integer (not merely a half-integer)
$$M=\sum_{{\mbox{\tiny thin wedges}}} k_{ab}.$$
This means that the sign of the $k_{ab}\Pi_{ab}$ term is actually irrelevant, and that the exponential of this term in the action is equal to $(-1)^M$.

\subsection{Asymptotic formula}

We can now state the asymptotic formula for the 4-simplex amplitude.  A formula is called asymptotic if the error term is bounded by a constant times one more power of $\lambda^{-1}$ than that stated in the asymptotic formula.

Given a set of non-degenerate boundary data, then in the limit
$\lambda \rightarrow \infty$, and for $p_{ab} =  \gamma k_{ab}$,
\begin{enumerate}
\item  If the boundary state is the Regge state of the boundary geometry of a  Lorentzian 4-simplex $\sigma$ we obtain:
\be
f_4
\sim (-1)^{\chi+M}
\left(\frac{1}{\lambda}\right)^{12}
\left[
 N_{| \sigma}  \exp \left( i \lambda  \gamma   \sum_{a<b}  k_{ab}   \Theta_{ab} \right)
 +
 N_{| {P \sigma}}  \exp \left( -i \lambda  \gamma   \sum_{a<b}  k_{ab}   \Theta_{ab} \right)
 \right].
\ee
$N_{| \sigma}$ and $N_{| {P \sigma}}$ are independent of $\lambda$ and are given below.

\item  If the boundary state is the Regge state of the boundary geometry of a 4d Euclidean 4-simplex $\sigma_E$ we obtain:
\be
f_4
\sim (-1)^{\chi}
\left(\frac{1}{\lambda}\right)^{12}
\left[
 N_{|\sigma_E}  \exp \left( {i} \lambda    \sum_{a<b}  k_{ab}   \Theta^E_{ab} \right)
 +
 N_{|P \sigma_E}  \exp \left( -{i} \lambda    \sum_{a<b}  k_{ab}   \Theta^E_{ab} \right)
 \right].
\ee
$\Theta^E_{ab}$ is the dihedral angle of the Euclidean 4-simplex.

\item If the boundary state is not that of the boundary of a non-degenerate 4-simplex but allows a single vector geometry $V$ as solution, then for an appropriate phase choice the asymptotic formula is:
\bea
f_4
&\sim&  \left(\frac{2 \pi}{\lambda}\right)^{12}
N_{| V}.
\eea
The number $N_{| V}$ is independent of $\lambda$.

\item  For a set of boundary data that is neither a non-degenerate Lorentzian 4-geometry nor admits a vector geometry solution,
the amplitude is suppressed for large
$\lambda$.
\be
f_4 = o(\lambda^{-K}) \quad\quad\quad\quad \forall \text{ non-negative integer } K
\ee
\end{enumerate}

The numerical factors $N$ of the stationary phase formula have to be evaluated at the critical points and are given by:

\bea
N_{|\mbox{\tiny crit}} &=&   (2\pi)^{22}        \frac{2^4}{    \sqrt{ \det H_{|\mbox{\tiny crit}} }} \prod_{a<b}   2k_{ab}   c_{ab}  {\Omega_{ab}}_{|\mbox{\tiny crit}}
\nn \\
&=&
 2^{36} \pi^{12}\left( \frac{1+i\gamma}{1-i\gamma}\right)^5 \frac{1}{\sqrt{ \det H_{|\mbox{\tiny crit}} }}
\prod_{a<b}  k_{ab} {\Omega_{ab}}_{|\mbox{\tiny crit}}
\eea

In the Lorentzian case the two contributions to the asymptotics correspond to the parity related reconstructions of this 4 simplex geometry, $\sigma$ and $P\sigma$. The factor $(2\pi)^{22} $ comes from the stationary phase formula as the integral has $6\times 4$ dimensions coming from the $\SLtwoC$ integrations and $20$ dimensions from the $z$ variables.  Since the formula is asymptotic, we have used $d_{\lambda k} \sim 2\lambda k$ and cancelled the scaling from the coefficients.  The additional factor $2^4$ comes from the fact that both spin lifts at the critical points give the same contribution to the action.  $H_{|\mbox{\tiny crit}}$ is the Hessian matrix of the action \eqref{staction} evaluated at the critical points; this is evaluated in appendix \ref{hessian section}.
The product $\prod{\Omega_{ab}}_{|\mbox{\tiny crit}}$ is the measure term evaluated at the critical points.  A choice of coordinate must be made to evaluate this. However the ratio of $\prod\Omega_{ab}$ with $\sqrt{ \det H }$ is independent of this choice of coordinates. The constant
$c_{ab}$ is equal to $\frac1\pi\sqrt{\frac{1+i\gamma}{1-i\gamma}}$.

For the Euclidean case we get contributions from the self dual and anti self dual part of the bivector geometry which combine to give the full 4-dimensional Euclidean bivector geometry of a 4-simplex $\sigma_E$. The phase part of the action for Euclidean boundary data is evaluated in \cite{su2paper}. The $N$ are the same as above but evaluated at the appropriate critical points.
The dihedral angle $\Theta^E_{ab}$ of a Euclidean 4-simplex arises in the following way.  For the case of Euclidean boundary data, there are two $\SU(2)$ solutions, say $X^+,X^-$, to the critical point equations \eqref{crit1}.  For these solutions the boost parameter $r_{ab}=0$ but the phase term $\theta_\pm$ remains.  The interpretation of the critical points is a pair of non-degenerate Euclidean 4-simplices $\sigma_E$ with $X^+$ as the selfdual part of its bivector geometry and $P \sigma_E$ with $X^-$ respectively. This is described in detail in \cite{Barrett:2009gg}. The Regge phase choice implies that $\theta_+ + \theta_- =0$ and $\theta_+ = \frac{1}{2}\Theta^E_{ab}$.  Combining this gives case 2. The Hessian and $\Omega_{ab}$ are evaluated on the elements $X^+ \in \SU(2) \subset \SLtwoC$ for the critical point corresponding to $\sigma_E$ and $X^-$ for $P \sigma_E$.

The case of a single vector geometry proceeds analogously. $N_{|V}$ is evaluated on the single $\SU(2)$ solution that defines the vector geometry $V$.
An appropriate phase choice for the boundary data such that the phase does not depend on $\lambda$ and the geometry of these solutions is described in \cite{su2paper}. In particular it was shown there that no such vector geometries exist for Lorentzian boundary data.

For the final case, no critical points exist and the stationary phase theorem tells us that the amplitude is suppressed.

By the classification of critical points in section \ref{class-solutions}  this concludes our asymptotic analysis of the amplitude.

\section{Conclusion}\label{conclusions}

In this work we have defined a graphical calculus for the unitary representations of Lorentz group, and used it to give a systematic definition of the 4-simplex amplitude in the case of Lorentzian quantum gravity. The asymptotic analysis of the amplitude has some surprising features. In the corresponding Euclidean quantum gravity problem analysed in \cite{Barrett:2009gg}, there was a puzzling superposition of terms with the Regge action multiplied by the Immirzi parameter and terms with the Regge action not multiplied by the Immirzi parameter.

In the Lorentzian quantum gravity analysed here, these two phenomena are separated out. The terms with the Immirzi parameter occur for boundary data of a Lorentzian metric and involve the Lorentzian Regge action. For this case, the result is much cleaner than for the Euclidean theory, as these are the only terms for this boundary data.
The terms without the Immirzi parameter occur for boundary data of a Euclidean metric and, rather surprisingly, involve the Euclidean Regge action, still in an oscillatory manner, and so are not suppressed. The physical significance of this is still unclear; the Euclidean action might be expected in relation to a tunnelling phenomenon, but then the amplitude would be exponentially damped, not oscillatory. Another surprise is that these terms are related to the $\SU(2)$ BF theory, and not to the Euclidean quantum gravity amplitudes. It is of course possible that in a state sum model, these terms are topological and do not contribute to the dynamics. This possibility is a topic for future work.

It is also important to mention the further case of non Regge-like boundary data. There are an extra five parameters for the boundary data in this case, so naively one might expect these to dominate. However this depends on further analysis of the phase factor for this case, which may turn out to be trivial.

Another feature which deserves further anaylsis is the formula for the Hessian. For example, we do not know yet whether the parity-related terms involving the Regge action and minus the Regge action occur with equal magnitude, or if one is heavily favoured over the other. Also, a geometric formula for the Hessian along the lines of the Ponzano-Regge formula is missing.

Another result of this work is that a condition for the existence of stationary points of the action is that $p_{ab} = \gamma k_{ab}$ for some constant $\gamma$.  This is exactly the same restriction on the representation labels derived in \cite{Engle:2007wy}, by different methods, where $\gamma$ is the Immirzi parameter.

Finally, it would be important to extend the results obtained here to give the asymptotics for the case of a triangulation of a manifold. This
was considered for the three-dimensional Ponzano-Regge model in \cite{Dowdall2010} and for the Euclidean signature four-dimensional models in
\cite{Conrady:2008mk}.

\section*{Acknowledgements} Eugenio Bianchi and Carlo Rovelli are thanked for useful discussions.
WF is supported by the Royal Commission for the Exhibition of
1851.  RD and FH are funded by EPSRC doctoral grants. RP thanks the QG research networking programme of the ESF for a visit grant.

\appendix

\section{The Hessian}
\label{hessian section}

Here we calculate the Hessian matrix required in the stationary phase formula.

The Hessian is defined as the matrix of second derivatives of the action where the variable $X_5$ has been gauge fixed to the identity.
We split the Hessian matrix into derivatives with respect to the $X_a$ variables and derivatives with respect to the $z_{ab}$.
The Hessian will then be a $44 \times 44$ matrix of the form
\be
H =
\left(
\begin{array}{ccc}
    H^{XX}          &    H^{X z}    \\
    H^{z X}         &    H^{z z}
\end{array}
\right) \ee
 We will now describe each block of this matrix.
 $H^{XX}$ is a $24 \times 24$ matrix containing only derivatives with respect to the $X_a$.  Note that due to the form of the action, derivatives with respect to two different variables will be zero and it will be block diagonal
\be
H^{X X} =
\left(
\begin{array}{cccc}
    H^{X_1 X_1}    & 0           &  0          & 0  \\
    0              & H^{X_2 X_2} &  0          & 0  \\
    0              & 0           & H^{X_3 X_3} & 0  \\
    0              & 0           &  0          &    H^{X_4 X_4}\\
\end{array}
\right)
\ee
Each $H^{X_1 X_1}$ is a $6 \times 6$ matrix.  The variation has been performed by splitting the $\SL(2,\C)$ element into a boost and a rotation generator.  This gives
\be
H^{X_a X_a} =
\left(
\begin{array}{cc}
    H^{X^R_a X^R_a}    & H^{X^B_a X^R_a}          \\
    H^{X^R_a X^B_a}    & H^{X^B_a X^B_a}     \\
\end{array}
\right)
\ee
With
\bea
H^{X^R_a X^R_a}_{ij}&=& \sum_{b \neq a} -2k_{ab}( \delta^{ij} - n_{\xi_{ab}}^i n_{\xi_{ab}}^j  )      \nn \\
H^{X^B_a X^R_a}_{ij}&=& i\sum_{b \neq a} 2k_{ab}( \delta^{ij} - n_{\xi_{ab}}^i n_{\xi_{ab}}^j  )       \nn \\
H^{X^R_a X^B_a}_{ij}&=& i\sum_{b \neq a} 2k_{ab}( \delta^{ij} - n_{{\xi_{ab}}}^i n_{\xi_{ab}}^j  )        \nn \\
H^{X^B_a X^B_a}_{ij}&=& \sum_{b \neq a}  -4i p_{ab}( \delta^{ij} - n_{{\xi_{ab}}}^i n_{{\xi_{ab}}}^j  )    -2k_{ab}( \delta^{ij} - n_{{\xi_{ab}}}^i n_{{\xi_{ab}}}^j  )        \nn \\
\eea
Where $i,j = 1,2,3$ and we have used the critical and stationary point equations.

$H^{z z}$ is a $20 \times 20$ matrix.  The derivatives are with respect to the spinor variables $z_{ab}$ on each of the ten edges $ab$ of the amplitude.
To perform these derivatives we must choose a section for $z_{ab}$.

Next the mixed spinor and $\SL(2,\C)$ derivatives.  We have arranged the derivatives in the order of the orientation $a<b$, ie $z_{12 },z_{13 },z_{14 },z_{15 },z_{23 },z_{24 },z_{25},z_{34 },z_{35 },z_{45 }$. The matrix $ H^{X \bar{z}}$ is a $24 \times 20 $ matrix with the following non-zero entries
\begin{multline}
H^{ X \bar{z} } = \nn \\
\left(
\begin{array}{cccccccccc}
H^{X_1 \bar{z}_{12}} & H^{X_1 \bar{z}_{13}} & H^{X_1 \bar{z}_{14}} & H^{X_1 \bar{z}_{15}} &      0           &  0      &  0     &       0       & 0 &0\\
H^{X_2 \bar{z}_{12}} & 0                & 0                & 0       & H^{X_2 \bar{z}_{23}} & H^{X_2 \bar{z}_{24}}& H^{X_2 \bar{z}_{25}} &   0   &0   & 0 \\
0                & H^{X_3 \bar{z}_{13}} & 0 & 0 & H^{X_2 \bar{z}_{23}}  &  0      &  0     &   H^{X_3 \bar{z}_{34}}    &H^{X_3 \bar{z}_{35}}& 0 \\
0                & 0 & H^{X_4 \bar{z}_{14}} & 0 &    0  & H^{X_4 \bar{z}_{24}} &  0     &  H^{X_4 \bar{z}_{34}} &0 &H^{X_4 \bar{z}_{45}} \\
\end{array}
\right)
\end{multline}
These derivatives also require a choice of section for $z_{ab}$.

\section{The Lorentz algebra}

In this Appendix, we summarise the conventions used throughout this paper regarding the Lorentz algebra.
The Lie algebra $\so(3,1)$ of the Lorentz group is a real, semi-simple Lie algebra of dimension six. A basis of $\so(3,1)$ is provided by the generators $(L_{\alpha \beta})_{\alpha<\beta = 0,...,3}$. The Lie algebra structure is coded in the brackets
\be
[L_{\alpha \beta}, L_{\gamma \delta} ] = - \eta_{\alpha \gamma} \, L_{\beta  \delta} + \eta_{\alpha \delta} \, L_{\beta \gamma} + \eta_{\beta \gamma} \, L_{\alpha \delta} - \eta_{\beta \delta} \, L_{\alpha \gamma}, \nn
\ee
where $\eta$ is the standard Minkowski metric with signature $-+++$.

It is convenient to decompose any (infinitesimal) Lorentz transformation into a purely spatial rotation and a hyperbolic rotation, or boost. This is achieved by introducing the rotation and boost generators respectively given by
\be
J_i = \frac{1}{2} \, \epsilon_{i}^{\;\, jk} L_{jk}, \;\;\;\; \mbox{and} \;\;\;\; K_i = L_{i0}, \;\;\;\;\;\;\;\; i,j,k = 1,2,3,
\ee
where $\epsilon_{ijk}$ is the three-dimensional Levi-Cevita tensor.

Using the Lie algebra structure of $\so(3,1)$ displayed above, it is immediate to check the following commutation relations between the rotation and boost generators
\be
[J_i, J_j] = - \epsilon_{ij}^{\;\; k} J_k, \;\;\;\;\;\; [J_i, K_j] = - \epsilon_{ij}^{\;\; k} K_k, \;\;\;\;\;\; [K_i, K_j] =  \epsilon_{ij}^{\;\; k} J_k.
\ee

The finite dimensional representations of the Lorentz algebra used in this paper are the spinor and vector representations. In the spinor representation $\rho : \so(3,1) \rightarrow \mathrm{End} \, \C^2$, the rotation and boost generators are given explicitly in terms of the Hermitian Pauli matrices
\be
\sigma_1 =
\left( \begin{array}{ll}
0 & 1 \\ \nn
1 & 0
       \end{array} \right) ,
\hspace{2mm}
\sigma_2 =
\left( \begin{array}{ll}
0 & -i \\ \nn
i & 0
       \end{array} \right) ,
\hspace{2mm}
\sigma_3 =
\left( \begin{array}{ll}
1 & 0 \\ \nn
0 & -1
       \end{array} \right) ,
\ee
\\
by the following expressions
\be
\rho(J_i) = \frac{i}{2} \sigma_i, \;\;\;\;\;\; \mbox{and} \;\;\;\;\;\; \rho(K_i) = \frac{1}{2} \sigma_i.
\ee
This is immediate to check by using the property $[\sigma_i,\sigma_j] = 2 i \epsilon_{ij}^{\;\; k} \sigma_k$ of the Pauli matrices.
Throughout the text, the map $\rho$ is kept implicit when there is no possible confusion.

In fact, the above presentation gives the explicit isomorphism between the Lorentz algebra $\so(3,1)$ and the realification $\sl(2,\C)_{\R}$ of the Lie algebra of the two-dimensional complex unimodular group $\SL(2,\C)$ because
\be
\sl(2,\C)_{\R} = (\su(2)^{\C})_{\R} \cong \su(2) \oplus i \su(2), \nn
\ee
where the direct sum is at the level of vector spaces.

Finally, we also used explicitly the vector representation $\pi : \so(3,1) \rightarrow \mathrm{End} \, \R^{3,1}$ of the Lorentz algebra in which the matrix elements of the $L_{\alpha \beta}$ generators are given by
\be
\pi(L_{\alpha \beta})^I_{\;J} = \delta^I_{\alpha} \eta_{\beta J} - \eta_{\alpha J} \delta^I_{\beta}, \;\;\;\;\;\;\;\; I,J = 0,...,3. \nn
\ee
From the above expression, it is immediate to compute the matrix elements of the image of the rotation and boost generators in the vector representation:
\be
\pi(J_i)^I_{\;J} = \epsilon_{i \; J}^{\;I}, \;\;\;\;\;\; \mbox{and} \;\;\;\;\;\; \pi(K_i)^I_{\;J} =  \delta^I_{i} \eta_{0 J} - \eta_{i J} \delta^I_{0}.
\ee
Therefore, an arbitrary element $L$ in the Lorentz algebra expressed in the rotation/boost basis as follows $L = \mathbf{r} \cdot \mathbf{J} + \mathbf{b} \cdot \mathbf{K}$ is given by the four-by-four matrix
\be
\label{vectorrep}
\pi(L) = \left[ \begin{array}{cccc} 0 & - b^1 & - b^2 & - b^3 \\ - b^1 & 0 & r^3 & - r^2 \\ - b^2 & - r^3 & 0 & r^1 \\ - b^3 & r^2 & -r^1 & 0 \end{array} \right],
\ee
in the vector representation.

\bibliographystyle{hieeetr}
\bibliography{Bibliography2009}

\end{document}

%% file: vertex.pstex_t
\begin{picture}(0,0)%
\includegraphics{vertex.pstex}%
\end{picture}%
\setlength{\unitlength}{3947sp}%
\begingroup\makeatletter\ifx\SetFigFont\undefined%
\gdef\SetFigFont#1#2#3#4#5{%
  \reset@font\fontsize{#1}{#2pt}%
  \fontfamily{#3}\fontseries{#4}\fontshape{#5}%
  \selectfont}%
\fi\endgroup%
\begin{picture}(4148,1569)(3368,-3073)
\put(3368,-2394){\makebox(0,0)[lb]{\smash{{\SetFigFont{12}{14.4}{\familydefault}{\mddefault}{\updefault}{\color[rgb]{0,0,0}$(k_1,p_1)$}%
}}}}
\put(4314,-1636){\makebox(0,0)[lb]{\smash{{\SetFigFont{12}{14.4}{\familydefault}{\mddefault}{\updefault}{\color[rgb]{0,0,0}$(k_2,p_2)$}%
}}}}
\put(5603,-1636){\makebox(0,0)[lb]{\smash{{\SetFigFont{12}{14.4}{\familydefault}{\mddefault}{\updefault}{\color[rgb]{0,0,0}$(k_3,p_3)$}%
}}}}
\put(6609,-2394){\makebox(0,0)[lb]{\smash{{\SetFigFont{12}{14.4}{\familydefault}{\mddefault}{\updefault}{\color[rgb]{0,0,0}$(k_4,p_4)$}%
}}}}
\end{picture}%

%% file: maximum.pstex_t
\begin{picture}(0,0)%
\includegraphics{maximum.pstex}%
\end{picture}%
\setlength{\unitlength}{3947sp}%
\begingroup\makeatletter\ifx\SetFigFont\undefined%
\gdef\SetFigFont#1#2#3#4#5{%
  \reset@font\fontsize{#1}{#2pt}%
  \fontfamily{#3}\fontseries{#4}\fontshape{#5}%
  \selectfont}%
\fi\endgroup%
\begin{picture}(2184,917)(3226,-1569)
\put(3226,-811){\makebox(0,0)[lb]{\smash{{\SetFigFont{12}{14.4}{\familydefault}{\mddefault}{\updefault}{\color[rgb]{0,0,0}$(k,p)$}%
}}}}
\end{picture}%

%% file: crossing.pstex_t
\begin{picture}(0,0)%
\includegraphics{crossing.pstex}%
\end{picture}%
\setlength{\unitlength}{3947sp}%
\begingroup\makeatletter\ifx\SetFigFont\undefined%
\gdef\SetFigFont#1#2#3#4#5{%
  \reset@font\fontsize{#1}{#2pt}%
  \fontfamily{#3}\fontseries{#4}\fontshape{#5}%
  \selectfont}%
\fi\endgroup%
\begin{picture}(2099,1224)(3226,-1573)
\put(3226,-736){\makebox(0,0)[lb]{\smash{{\SetFigFont{12}{14.4}{\familydefault}{\mddefault}{\updefault}{\color[rgb]{0,0,0}$(k,p)$}%
}}}}
\put(4726,-736){\makebox(0,0)[lb]{\smash{{\SetFigFont{12}{14.4}{\familydefault}{\mddefault}{\updefault}{\color[rgb]{0,0,0}$(k',p')$}%
}}}}
\end{picture}%

%% file: RImove.pstex_t
\begin{picture}(0,0)%
\includegraphics{RImove.pstex}%
\end{picture}%
\setlength{\unitlength}{3947sp}%
\begingroup\makeatletter\ifx\SetFigFont\undefined%
\gdef\SetFigFont#1#2#3#4#5{%
  \reset@font\fontsize{#1}{#2pt}%
  \fontfamily{#3}\fontseries{#4}\fontshape{#5}%
  \selectfont}%
\fi\endgroup%
\begin{picture}(4213,1821)(3226,-1573)
\put(3226,-736){\makebox(0,0)[lb]{\smash{{\SetFigFont{12}{14.4}{\familydefault}{\mddefault}{\updefault}{\color[rgb]{0,0,0}$(k,p)$}%
}}}}
\put(5401,-1111){\makebox(0,0)[lb]{\smash{{\SetFigFont{12}{14.4}{\familydefault}{\mddefault}{\updefault}{\color[rgb]{0,0,0}=}%
}}}}
\put(6901,-886){\makebox(0,0)[lb]{\smash{{\SetFigFont{12}{14.4}{\familydefault}{\mddefault}{\updefault}{\color[rgb]{0,0,0}$(k,p)$}%
}}}}
\end{picture}%

%% file: spinnetwork.pstex_t
\begin{picture}(0,0)%
\includegraphics{spinnetwork.pstex}%
\end{picture}%
\setlength{\unitlength}{3947sp}%
\begingroup\makeatletter\ifx\SetFigFont\undefined%
\gdef\SetFigFont#1#2#3#4#5{%
  \reset@font\fontsize{#1}{#2pt}%
  \fontfamily{#3}\fontseries{#4}\fontshape{#5}%
  \selectfont}%
\fi\endgroup%
\begin{picture}(3624,3852)(2989,-3106)
\put(3301,614){\makebox(0,0)[lb]{\smash{{\SetFigFont{12}{14.4}{\familydefault}{\mddefault}{\updefault}{\color[rgb]{0,0,0}$(k_1,p_1)$}%
}}}}
\put(4276, 89){\makebox(0,0)[lb]{\smash{{\SetFigFont{12}{14.4}{\familydefault}{\mddefault}{\updefault}{\color[rgb]{0,0,0}$(k_2,p_2)$}%
}}}}
\put(4426,-811){\makebox(0,0)[lb]{\smash{{\SetFigFont{12}{14.4}{\familydefault}{\mddefault}{\updefault}{\color[rgb]{0,0,0}$(k_3,p_3)$}%
}}}}
\put(3526,-3061){\makebox(0,0)[lb]{\smash{{\SetFigFont{12}{14.4}{\familydefault}{\mddefault}{\updefault}{\color[rgb]{0,0,0}$\tau_1$}%
}}}}
\put(5926,-3061){\makebox(0,0)[lb]{\smash{{\SetFigFont{12}{14.4}{\familydefault}{\mddefault}{\updefault}{\color[rgb]{0,0,0}$\tau_2$}%
}}}}
\end{picture}%

%% file: spinnetwork2.pstex_t
\begin{picture}(0,0)%
\includegraphics{spinnetwork2.pstex}%
\end{picture}%
\setlength{\unitlength}{3947sp}%
\begingroup\makeatletter\ifx\SetFigFont\undefined%
\gdef\SetFigFont#1#2#3#4#5{%
  \reset@font\fontsize{#1}{#2pt}%
  \fontfamily{#3}\fontseries{#4}\fontshape{#5}%
  \selectfont}%
\fi\endgroup%
\begin{picture}(5632,3027)(2026,-3106)
\put(2026,-1336){\makebox(0,0)[lb]{\smash{{\SetFigFont{12}{14.4}{\familydefault}{\mddefault}{\updefault}{\color[rgb]{0,0,0}$(k_2,p_2)$}%
}}}}
\put(3526,-3061){\makebox(0,0)[lb]{\smash{{\SetFigFont{12}{14.4}{\familydefault}{\mddefault}{\updefault}{\color[rgb]{0,0,0}$\tau_1$}%
}}}}
\put(4276,-211){\makebox(0,0)[lb]{\smash{{\SetFigFont{12}{14.4}{\familydefault}{\mddefault}{\updefault}{\color[rgb]{0,0,0}$(k_3,p_3)$}%
}}}}
\put(6751,-961){\makebox(0,0)[lb]{\smash{{\SetFigFont{12}{14.4}{\familydefault}{\mddefault}{\updefault}{\color[rgb]{0,0,0}$(k_1,p_1)$}%
}}}}
\put(5926,-3061){\makebox(0,0)[lb]{\smash{{\SetFigFont{12}{14.4}{\familydefault}{\mddefault}{\updefault}{\color[rgb]{0,0,0}$\tau_2$}%
}}}}
\end{picture}%

%% file: vertexperm.pstex_t
\begin{picture}(0,0)%
\includegraphics{vertexperm.pstex}%
\end{picture}%
\setlength{\unitlength}{3947sp}%
\begingroup\makeatletter\ifx\SetFigFont\undefined%
\gdef\SetFigFont#1#2#3#4#5{%
  \reset@font\fontsize{#1}{#2pt}%
  \fontfamily{#3}\fontseries{#4}\fontshape{#5}%
  \selectfont}%
\fi\endgroup%
\begin{picture}(6624,2157)(2389,-3106)
\put(5776,-1861){\makebox(0,0)[lb]{\smash{{\SetFigFont{12}{14.4}{\familydefault}{\mddefault}{\updefault}{\color[rgb]{0,0,0}=}%
}}}}
\put(2926,-3061){\makebox(0,0)[lb]{\smash{{\SetFigFont{12}{14.4}{\familydefault}{\mddefault}{\updefault}{\color[rgb]{0,0,0}$\tau_1$}%
}}}}
\put(4426,-3061){\makebox(0,0)[lb]{\smash{{\SetFigFont{12}{14.4}{\familydefault}{\mddefault}{\updefault}{\color[rgb]{0,0,0}$\tau_2$}%
}}}}
\put(6826,-3061){\makebox(0,0)[lb]{\smash{{\SetFigFont{12}{14.4}{\familydefault}{\mddefault}{\updefault}{\color[rgb]{0,0,0}$\tau_2$}%
}}}}
\put(8326,-3061){\makebox(0,0)[lb]{\smash{{\SetFigFont{12}{14.4}{\familydefault}{\mddefault}{\updefault}{\color[rgb]{0,0,0}$\tau_1$}%
}}}}
\end{picture}%